\documentclass[
reprint, 
superscriptaddress, 
showpacs,
 amsmath,amssymb,
 aps,
prl
]{revtex4-1}

\usepackage{graphicx}
\usepackage{dcolumn}
\usepackage{bm}
\usepackage{appendix}
\usepackage{hyperref}
\usepackage[mathlines]{lineno}
\usepackage[marginal]{footmisc} 
\usepackage{float}

\begin{document}

\preprint{APS/123-QED}

\title{Observation of Anomalous Information Scrambling in a Rydberg Atom Array}

\author{Xinhui Liang}
\thanks{These authors contributed equally to this work.}
\affiliation{State Key Laboratory of Low Dimensional Quantum Physics, Department of Physics, Tsinghua University, Beijing 100084, China}
\affiliation{Beijing Academy of Quantum Information Sciences, Beijing 100193, China}

\author{Zongpei Yue}
\thanks{These authors contributed equally to this work.}
\affiliation{State Key Laboratory of Low Dimensional Quantum Physics, Department of Physics, Tsinghua University, Beijing 100084, China}
\affiliation{Frontier Science Center for Quantum Information, Beijing 100084, China}

\author{Yu-Xin Chao}
\affiliation{State Key Laboratory of Low Dimensional Quantum Physics, Department of Physics, Tsinghua University, Beijing 100084, China}

\author{Zhen-Xing Hua}
\affiliation{State Key Laboratory of Low Dimensional Quantum Physics, Department of Physics, Tsinghua University, Beijing 100084, China}

\author{Yige Lin}
\affiliation{Division of Time and Frequency Metrology, National Institute of Metrology, Beijing 100029, China}

\author{Meng Khoon Tey}
\email{mengkhoon_tey@mail.tsinghua.edu.cn}
\affiliation{State Key Laboratory of Low Dimensional Quantum Physics, Department of Physics, Tsinghua University, Beijing 100084, China}
\affiliation{Frontier Science Center for Quantum Information, Beijing 100084, China}
\affiliation{Hefei National Laboratory, Hefei, Anhui 230088, China}

\author{Li You}
\email{lyou@mail.tsinghua.edu.cn}
\affiliation{State Key Laboratory of Low Dimensional Quantum Physics, Department of Physics, Tsinghua University, Beijing 100084, China}
\affiliation{Beijing Academy of Quantum Information Sciences, Beijing 100193, China}
\affiliation{Frontier Science Center for Quantum Information, Beijing 100084, China}
\affiliation{Hefei National Laboratory, Hefei, Anhui 230088, China}

\date{July 31, 2025}

\begin{abstract}
Quantum information scrambling, which describes the propagation and effective loss of local information, is crucial for understanding the dynamics of quantum many-body systems. 
We report the observation of anomalous information scrambling in an atomic tweezer array with dominant van der Waals interaction. 
We characterize information spreading by an out-of-time-order correlator and observe persistent oscillations inside a suppressed linear light cone for the initial Néel state.
Such an anomalous dynamic, which differs from both generic thermal and many-body localized scenarios, originates from weak ergodicity breaking 
in quantum many-body scarred systems. 
\end{abstract}

\maketitle

\paragraph{Introduction—}
Quantum information scrambling in closed many-body systems has emerged as an effective method for exploring far-from-equilibrium physics. Characterized by the dispersal of local information to the entire system through many-body entanglement and correlation, this process is key to understanding the dynamics of thermalization and the evolution toward equilibrium~\cite{Swingle2018, LewisSwan2019, Xu2024Scrambling}. In general, a typical interacting system thermalizes under time evolution, leading to the emergence of ergodicity and linear light cones of information scrambling~\cite{rigol2008thermalization}. In contrast, a many-body localized system possesses an extensive number of conserved quantities that prevent the system from thermalization~\cite{nandkishore2015many,Abanin2019Colloquium}, resulting in full ergodicity breaking and a logarithmic light cone~\cite{Deng2017Logarithmic, fan2017out,huang2017out}. Quantum many-body scar~\cite{Bernien2017Probing,Bluvstein2021Controlling, Turner2018weak, Serbyn2021, Zhang2022Many, Su2023Observation,  Bies2001Scarring}, as another notable example deviating from the nominal understanding of ergodicity in the eigenstate thermalization hypothesis  (ETH)~\cite{Deutsch1991Quantum,Srednicki1994Chaos}, refers to a special class of eigenstates embedded in the sea of thermalizing eigenstates that partially break ergodicity~\cite{Turner2018weak, Serbyn2021}. Such systems were predicted to exhibit anomalous information scrambling characterized by persistent oscillations within suppressed light cones~\cite{Yuan2022Quantum}, a phenomenon which has not been observed until this work.

A powerful metric for quantifying the intricate scrambling of local information is the out-of-time-order correlator (OTOC)~\cite{Swingle2018, LewisSwan2019, Xu2024Scrambling}. We will focus on the form of the squared commutator
\begin{eqnarray}
C(t) = \langle\psi|[W(t),V]^{\dagger}[W(t),V]|\psi\rangle,
\label{eq:def2}
\end{eqnarray}
which measures how fast the noncommutativity between two 
quantum operations is established, where $W(t) = e^{i\mathcal{H}t}We^{-i\mathcal{H}t}$ represents the operator's evolution in the Heisenberg picture.

Experimental studies of the OTOC have been carried out in various physical systems, from nuclear spins~\cite{Li17Measuring, Wei18Exploring,Nie20Experimental, Niknam20Sensitivity}, trapped ions~\cite{Martin17Measuring, Landsman2019, Manoj20Quantum, Alaina22Experimental, Tian2022Testing}, and superconducting circuits~\cite{Xiao21Information, Jochen22Probing, Zhao22Probing, Wang22Information}, to nitrogen vacancy centers~\cite{Chen2020Detecting}, etc. A primary challenge in measuring the OTOC is the implementation of time-reversed dynamics. While allowed in principle for unitary evolutions, it is inherently difficult to achieve in practice, mainly due to large system sizes, limited experimental control capability of the many-body interactions, and inevitable experimental noises.
A variety of approaches are adopted to realize time-reversed evolution, employing digital simulation with quantum gates and the Suzuki-Trotter decomposition~\cite{Li17Measuring, Nie20Experimental, Wei18Exploring, Niknam20Sensitivity, Landsman2019,Xiao21Information, Alaina22Experimental} or fully analog simulation with controllable interactions~\cite{Martin17Measuring, Jochen22Probing, Zhao22Probing, Wang22Information}, etc. 
Randomized measurements~\cite{Manoj20Quantum, NieQuantum, Vermersch2019Probing} and several other protocols~\cite{Blocher2022Measuring, KastnerQuantum} can also be used to circumvent the time-reversed evolution. 

Recent advancements in atomic tweezer arrays offer a promising quantum information processing platform with single-qubit addressability and the potential to scale up to large sizes and programmable configurations~\cite{Saffman2010Quantum, Saffman2016Quantum, Schauss2018Quantum, Adams2019Rydberg, Browaeys2020Many, Wu2021A, Morgado2021Quantum, Manetsch2024A, Ma2022Universal, Anand2024A, Sheng2022Defect, Zhao2023Floquet, Gyger2024Continuous}. 
The strong van der Waals interactions, predominant in such systems, can give rise to Rydberg blockade and rich quantum phases~\cite{Bernien2017Probing, Semeghini2021Probing, Kim2024Realization}.
The presence of local kinematic constraints is also believed to give rise to quantum many-body scars.
By working in the Rydberg blockade regime where van der Waals interaction is the dominant energy scale, the system simplifies into a kinetically constrained model~\cite{Turner2018weak}.
The challenge of reversing the interaction sign can be bypassed in the truncated Hilbert space, and one is able to implement and measure the OTOC and observe the anomalous information scrambling.

\begin{figure*}[t]
\centering
\includegraphics[width=\linewidth]{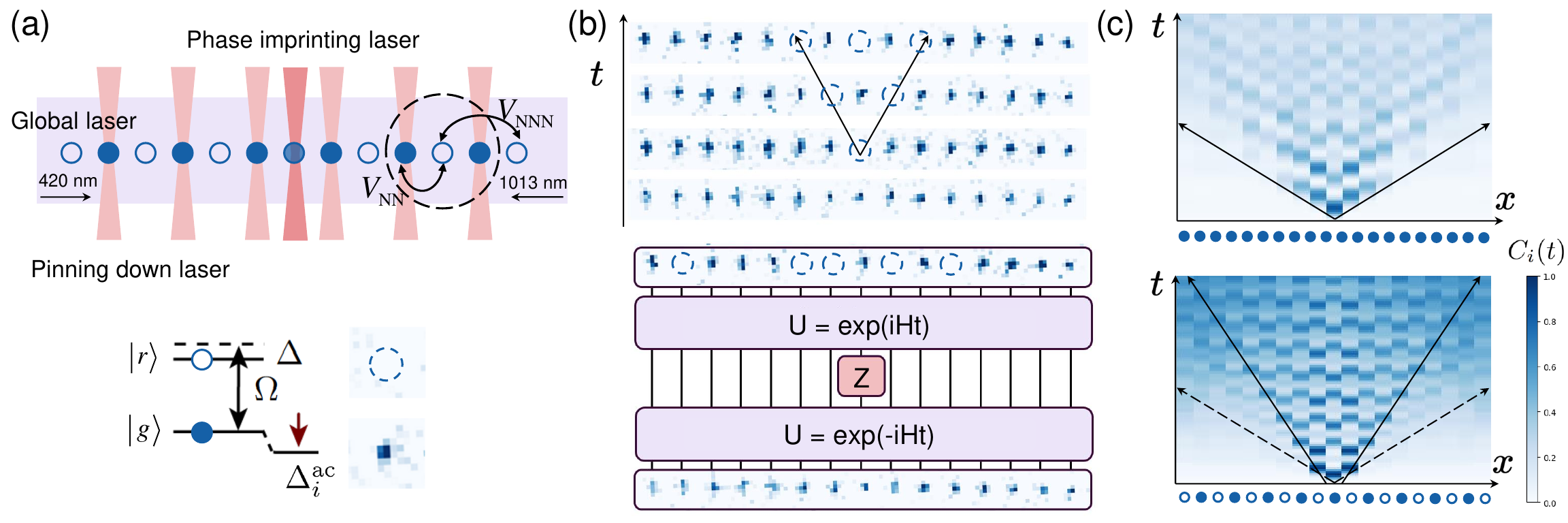}
\caption{\label{fig:demo}The illustration for observing the OTOC in
a one dimensional (1D) atom array. (a) The experimental setup consists of a 1D defect-free atom array. 
Qubits are encoded into atomic ground (solid circle) and Rydberg (empty circle) states. The evolution is driven by counterpropagating lasers with two-photon Rabi frequency $\Omega$ and detuning $\Delta$, 
while additional programmable laser beams generate single-qubit ac-Stark shift $\Delta_i^{{\rm ac}}$ used for addressable qubit rotation (phase imprinting laser in dark red) and select initial state preparation (pinning down laser in light red). The state readout is through destructive imaging, wherein the ground state is recaptured while the Rydberg state is irreversibly lost.
The relationship for the nearest-neighbor (NN) blockade strength $V_{\rm{NN}}\gg \Omega \gg V_{\rm{NNN}}$ constrains many-body dynamics to a subspace where two nearby atoms cannot be simultaneously in Rydberg states. (b) The experimental sequence to measure the OTOC. Local disturbance in the form of a single-qubit rotation $\sigma_z$ is applied to the central atom of the chain between the time-forward and -reversed evolutions. Information scrambling characterized by the ZZ-OTOC is probed by counting the difference of local Rydberg atom density between the initial and final states for varying evolution time $2t$. (c) Numerically calculated $C_{i}(t)$ for the two initial states: $\left|\bold{g}\right>=\left|g,g,...,g\right>$, show a linear light cone and the OTOC contrast decays fast inside the light cone, while the $Z_2$ Néel states $\left|{\bf Z}_2\right>=\left|r,g,r,g,...,r,g,r\right>$, possessing a large overlap with the scarred subspace, exhibits slower information spreading with persistent oscillations inside the light cone.
}
\end{figure*}

In this Letter, we report the observation of anomalous information scrambling in a Rydberg atom array. Our results show that the ZZ-OTOC ($W=\sigma_z^{(i)},V=\sigma_z^{(j)}$) for the initial $Z_2$ Néel state exhibits smaller butterfly velocity, a characteristic speed of quantum information spreading, compared to the polarized state with all atoms in the ground state. 
Within the linear light cone, the honeycomblike 
persistent oscillations, where the OTOC peaks at the same time, are observed.
Analogous features are also observed in the spatial Rydberg atom density and two-body correlation in time-forward dynamics.

\paragraph{Experimental setup—} 
Our experimental setup is illustrated in Fig.~\ref{fig:demo}(a). Neutral $^{87}$Rb atoms are 
rearranged into a linear single-atom array with equal spacing. They are initialized in the electronic ground state $\left| g \right \rangle = \left | 5S_{1/2}, F = 2,m_F = 2  \right \rangle$ and coupled to the Rydberg state $\left| r \right \rangle = \left | 70S_{1/2},J=1/2, m_J = 1/2 \right \rangle$ via the intermediate state $\left | 6P_{3/2}, F = 3,m_F = 3 \right \rangle$ using two counterpropagating lasers at 420 and 1013 nm, respectively, with effective two-photon Rabi frequency $\Omega$ and the bare two-photon detuning $\Delta$. When excited into the Rydberg state, neighboring atoms interact via a strong, repulsive van der Waals interaction $V(\mathbf{r}_{ij})$, which decays by the sixth power of interatomic separation $\mathbf{r}_{ij}$. The many-body Hamiltonian ($\hbar=1$) is approximately described by
\begin{equation}
    H=\frac{\Omega}2\sum_i{\sigma}_x^{(i)}-\sum_i(\Delta-\Delta_i^{{\rm ac}})n_i^r+\sum_{i<j}V(\mathbf{r}_{ij}){n_i^r}{n_j^r}
    \label{eq:rydberg},
\end{equation}
with $i$ indexing atomic site in the array, $n_i^r=({1+\sigma_{z}^{(i)}})/{2}$ is the on-site Rydberg atom density operator, and $\Delta_i^{{\rm ac}}$ represents the ac-Stark shift by individual addressing lasers.

In the limit of perfect NN blockade and negligible interactions for next-nearest neighbors (NNN) and beyond, or $V_{\rm{NN}}\gg \Omega \gg V_{\rm{NNN}}$, the many-body dynamics are governed by the PXP model
\begin{eqnarray}
H=\frac{\Omega}2\sum_{i=1}^{L-1}P_{i-1}\sigma_{x}^{(i)}P_{i+1}
\label{eq:pxp},
\end{eqnarray}
where $P_i=(1-\sigma_{z}^{(i)})/2$ is the ground state projection operator, and two-photon detuning $\Delta-\Delta_i^{{\rm ac}}$ is set to 0. This model constrains an atom from getting exited to the Rydberg state or deexcited back to the ground state if either of its nearest neighbors are in the Rydberg states. Although the system remains non-integrable and its dynamics chaotic, the PXP Hamiltonian possesses a small fraction of scarred eigenstates that violate the ETH~\cite{Serbyn2021}, manifested by periodic recurrences in the dynamics of entanglement entropy and local correlation function~\cite{Turner2018weak}. The $Z_2$ N\'{e}el state $\left|{\bf Z}_2\right>=\left|rgrg,...,rgr\right>$ is of particular interest, as it can be directly prepared and has a large overlap with the scarred subspace, making the PXP a pivotal model for exploring information scrambling in a constrained system.

\begin{figure}[t]
\centering
\includegraphics[width=\linewidth]{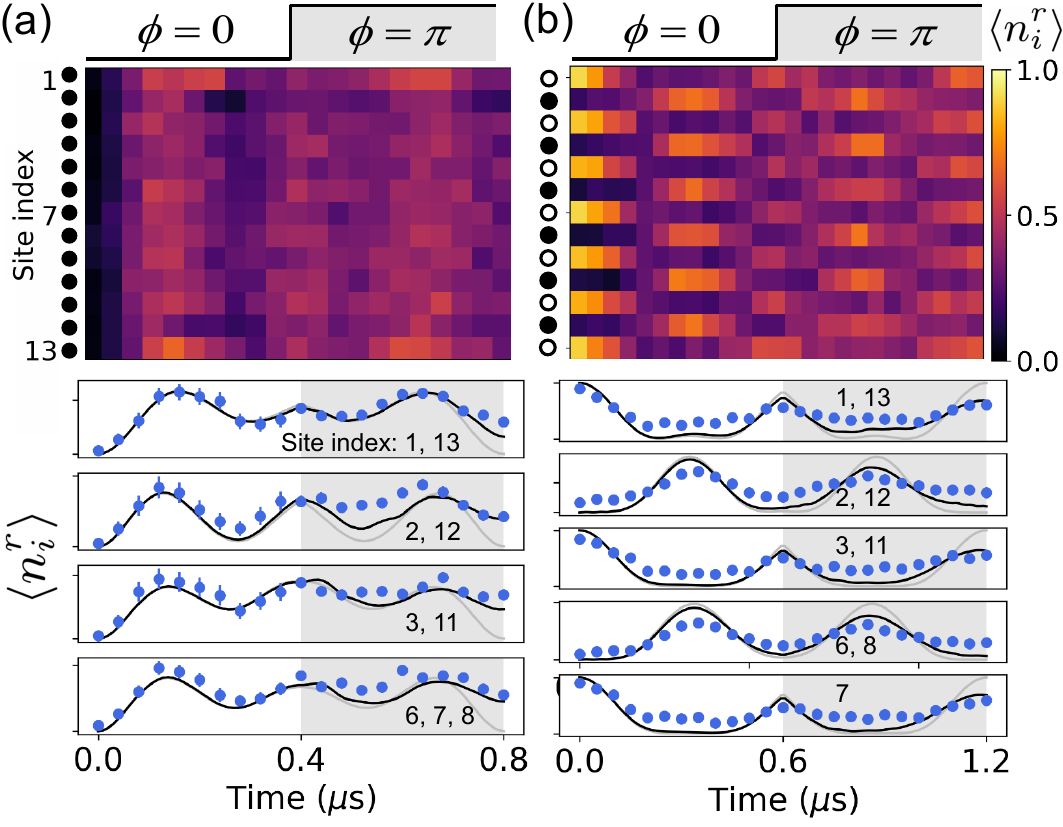}
\caption{\label{fig:timereverse}Time-reversed evolution is implemented through a $\pi$ phase jump of the global Rydberg laser, which changes the sign of the effective PXP Hamiltonian. The evolution starts from the state $\left|\bold{g}\right>=\left|gg,...,g\right>$ in (a) and the $\left|{\bf Z}_2\right>=\left|rgrg,...,rgr\right>$ Néel state in (b). The white and gray backgrounds in the headers indicate the phases before and after the phase jump, respectively. The upper panel displays the averaged Rydberg atom density at each site, while the lower panel provides a detailed comparison of the experimental results (symbols) with numerical simulations based on the actual Hamiltonian Eq. (\ref{eq:rydberg}) (black lines) and the PXP model Eq. (\ref{eq:pxp}) (gray lines). The deviations between the experimental data and the perfect time-reversed evolution of the PXP model primarily stem from the finite NNN interactions in the actual Hamiltonian.
The contrast in the $|{\bf Z}_2\rangle$ Néel state evolution is reduced due to imperfect preparation of the initial state.
}
\end{figure}

\paragraph{Measuring the ZZ-OTOC—}

The measurement of the OTOC for arbitrary initial states is time-consuming and challenging. In the work reported here, we simplify the task by 
selecting eigenstates of the operator $V=\sigma_z^{(i)}$,
as detailed in Sec. II of Supplemental Material~\cite{SM}. Although such a choice limits the number of measurable initial states, it does not affect our conclusions. 
We take the following two initial states: the $\left|{\bf Z}_2\right>$ 
and the fully polarized state $\left|\bold{g}\right>=\left|gg,...,g\right>$. 
The former represents a case with significant overlap to the scar subspace, while the latter is mainly outside the scar subspace. Further details can be found in Sec. VII of Supplemental Material~\cite{SM}.

We measure the OTOC through the Loschmidt echo protocol~\cite{Quan2006Decay, Macr2016Loschmidt, Jochen22Probing} as shown in Fig.~\ref{fig:demo}(b) where a precisely prepared initial state first undergoes time-forward evolution and then time-reversed evolutions designed to restore the system to its initial configuration. In between the time-forward and time-reversed evolution, a local operator $\sigma_z$
is applied to the central or the seventh atom of the 13 atom chain, disrupting the full reversibility. The choice of a local perturbation $W=\sigma_z^{(\text{7})}$ and measurement $V=\sigma_z^{(i)}$ provides the spatial resolution of information scrambling. The measured ZZ-OTOC thus simplifies to
$C_{i}(t)= \langle\psi|[{\sigma}_z^{(\text{7})}(t), {\sigma}_z^{(i)}]^{\dagger}[{\sigma}_z^{(\text{7})}(t), {\sigma}_z^{(i)}]|\psi\rangle/4\nonumber=\lvert \langle n_i^r(2t)\rangle - \langle n_i^r(0)\rangle \rvert$, 
where $\langle n^r_i(2t)\rangle$
represents the probability the $i$th atom in the Rydberg state after completing a Loschmidt echo with total evolution time of $2t$.
A nonzero valued $C_{i}(t)$ indicates the existence of finite information transport from the middle of the array to other sites. Dividing by 4 gives the convenient normalization, which reduces the OTOC value to counting the density of Rydberg atoms. By adjusting the echo duration $2t$ and the site index of the measurement operator, we obtain time- and site-resolved ZZ-OTOC.

\paragraph{Time-reversed evolution—}

The time-reversed evolution is implemented by imprinting a global $\pi$ phase jump to the Rydberg coupling laser (see Supplemental Material Sec. III~\cite{SM}), which introduces a negative sign to $\Omega$ and consequently to the overall PXP Hamiltonian.
To verify the efficacy of this solution, we measure the Loschmidt echo with two representative initial states $\left|\bold{g}\right>$
and $|{\bf Z}_2\rangle$.
As shown in Fig. \ref{fig:timereverse}, the averaged Rydberg atom occupation is found to be approximately symmetric with respect to the phase-jumping instant, supporting the understanding that the effective Hamiltonian before and after the jump is approximately opposite in sign.

Infidelity for the time-reversed evolution arises primarily from the ignored NNN interactions in the PXP model. An extra detuning at approximately twice the interaction strength $V_{\rm NNN}$ provides partial mitigation (see \cite{Yuan2022Quantum} and Supplemental Material Sec. V~\cite{SM}).

\paragraph{Emergent light cone with initial state $\left|\bold{g}\right>$—}

\begin{figure}[H]
\centering
\includegraphics[width=\linewidth]{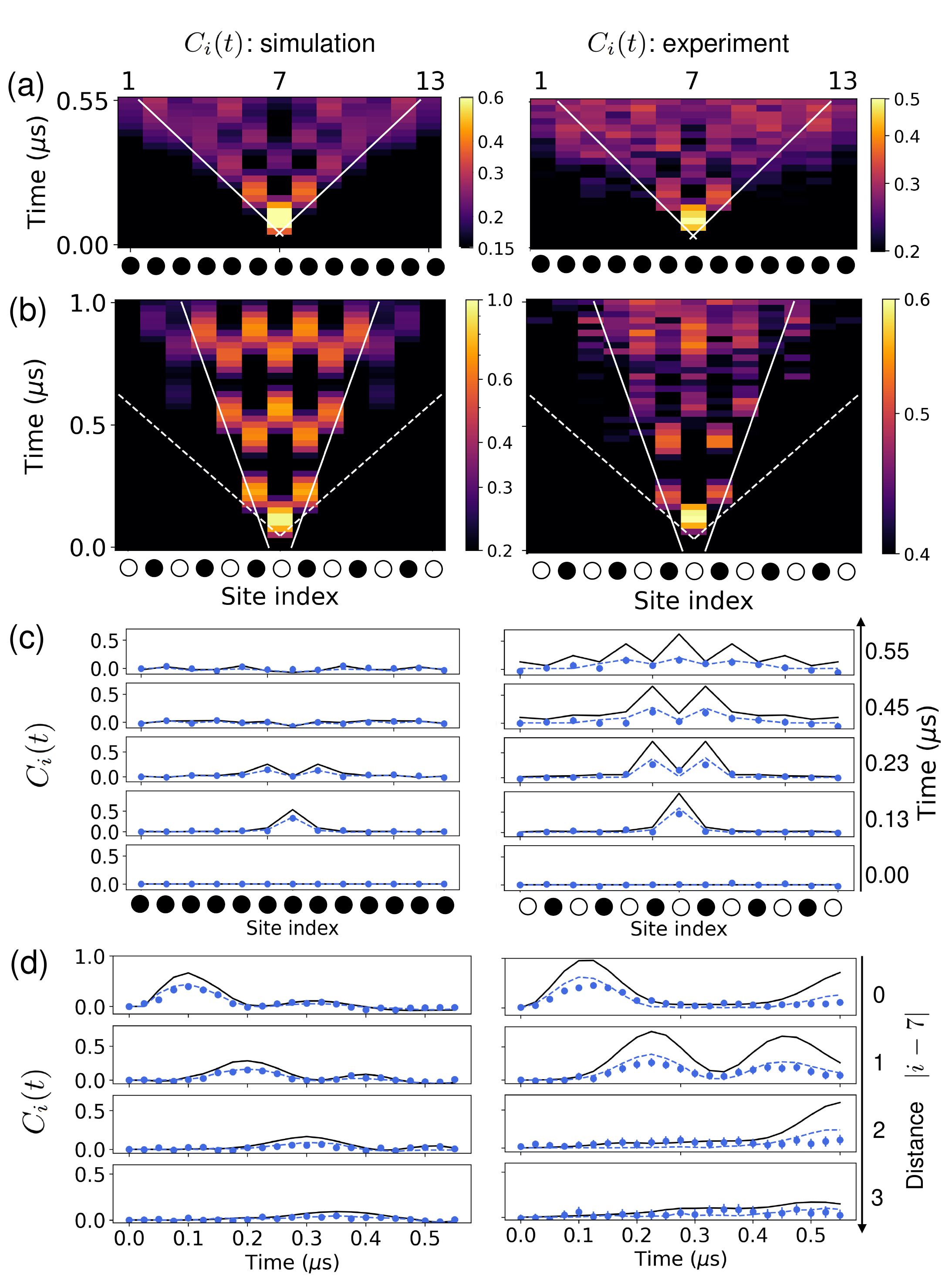}
\caption{\label{fig:otoc_combine}The observed ZZ-OTOC compared to numerical simulations. (a) and (b) are the $\left|\bold{g}\right>$ and $|{\bf Z}_2\rangle$ initial states, respectively.
The left and right panels in (a) and (b) denote respectively numerical and experimental results. A linear light-cone-like operator spreading is observed in both cases. The $Z_2$ Néel state shows a slower but highly synchronized pattern within the light cone, where ZZ-OTOC at different sites peak at the same time. Solid (dashed) lines serve as guides to compare with the light cones of the two different initial states. (c) Data at peak time of the periodic oscillations to emphasize the persistent and synchronized feature of the OTOC within the light cone for the $Z_2$ Néel state. (d) The spreading of the ZZ-OTOC from the addressed central spin (site 7) to other spins, averaged over data for specific spin $i$ ordered by the separation $|i-\text{7}|$. Experimental (symbols) and two simulated results are compared in the same subfigures: one using the actual Hamiltonian Eq. (\ref{eq:rydberg}) (solid lines) and the other with modeled decoherence sources from fluctuations in atomic positions and measured single-atom decoherence in experiment (dashed lines). Background subtraction is applied (see Supplemental Material Sec. V~\cite{SM}) 
in (c) and (d) to mitigate nonzero bias caused by imperfections in time-reversed evolution.
}
\end{figure}

The experimental and numerical results for the ZZ-OTOC are presented in Fig.~\ref{fig:otoc_combine}(a) for the initial state $\left|\bold{g}\right>$. A linear light-cone-like pattern of operator spreading is clearly visible, with butterfly velocity $v$ of 11.7 ± 0.5 sites per microsecond based on a linear fit (see Supplemental Material Sec. VI). The dimensionless butterfly velocity $2\pi v/\Omega$, which is normalized to the Rabi frequency $\Omega$, is 4.7 ± 0.2. It aligns well with the numerical result of 4.9 and represents a typical speed of information spreading.

Compared to numerical ones, the observed contrast is found lower. Such a degradation can be attributed primarily to decoherence that arises from three sources: variations in the atomic positions within the chain, which cause phase imprinting errors, the consequent fluctuations in the van der Waals interaction $V({\bf r}_{ij})$, and single-atom decoherence. 
Our experimental data compare well to numerical calculations shown in blue dashed lines after considering these decoherence effects, as detailed in Supplemental Material Sec. IV~\cite{SM}. This agreement implicates that the dominant impact of decoherence is reduction in signal-to-noise ratio of ZZ-OTOC. The fundamental scrambling dynamics, as characterized by the butterfly velocity, remains largely unaffected by the decoherence sources present in our system.

\paragraph{Anomalous information scrambling with initial 
Néel state—}

For the second set of experiments, the $Z_2$ Néel state is prepared 
by adiabatically driving the atoms from $\left|g\right>$ to $\left|r\right>$ globally, while inhibiting such a transfer for the odd sites by employing additional addressing lasers as shown in Fig.~\ref{fig:demo} (see Supplemental Material Sec. III~\cite{SM}).

The measurement results are presented in Fig. \ref{fig:otoc_combine}. Besides a slower butterfly velocity,
we observe an oscillatory operator spreading within the light cone, in stark contrast to the initial state $\left|\bold{g}\right>$. While slower operator spreading was also found in other ergodicity-breaking systems, such as the logarithmic-shaped light cone in systems with many-body localization~\cite{Abanin2019Colloquium, Deng2017Logarithmic, fan2017out, huang2017out}, the light cone we observe for the $|{\bf Z}_2\rangle$ initial state remains linear with a fitted 
dimensionless
butterfly velocity
of $1.4 \pm 0.3$, or in absolute value of $3.5 \pm 0.7$ sites per microsecond.
The characteristic periodic recurrences or ``synchronized'' oscillations within the light cone, marking the same time when the ZZ-OTOC peak across different sites, recur every 0.32 $\mu s$, which also aligns well with the revival period in the time-forward evolution of the $|{\bf Z}_2\rangle$ state. 
Experimental imperfections in the initial state preparation and time-reversed evolution reduce the long time OTOC amplitude, as discussed in SM Sec. IV~\cite{SM}, yet long-lasting periodic revivals remain visible in comparison to the $\left|\bold{g}\right>$ state under the same experimental conditions. 

\paragraph{Local Rydberg atom density and two-body correlation—}

To affirm the observed slower information spreading and the 
persistent
honeycomblike structure within the ZZ-OTOC light cone, we study the time-forward evolution of the $|{\bf Z}_2\rangle$ state and analyze directly its local Rydberg atom density and the propagation of two-body correlation~\cite{Surace2020Lattice}.

\begin{figure}[t]
\centering
\includegraphics[width=\linewidth]{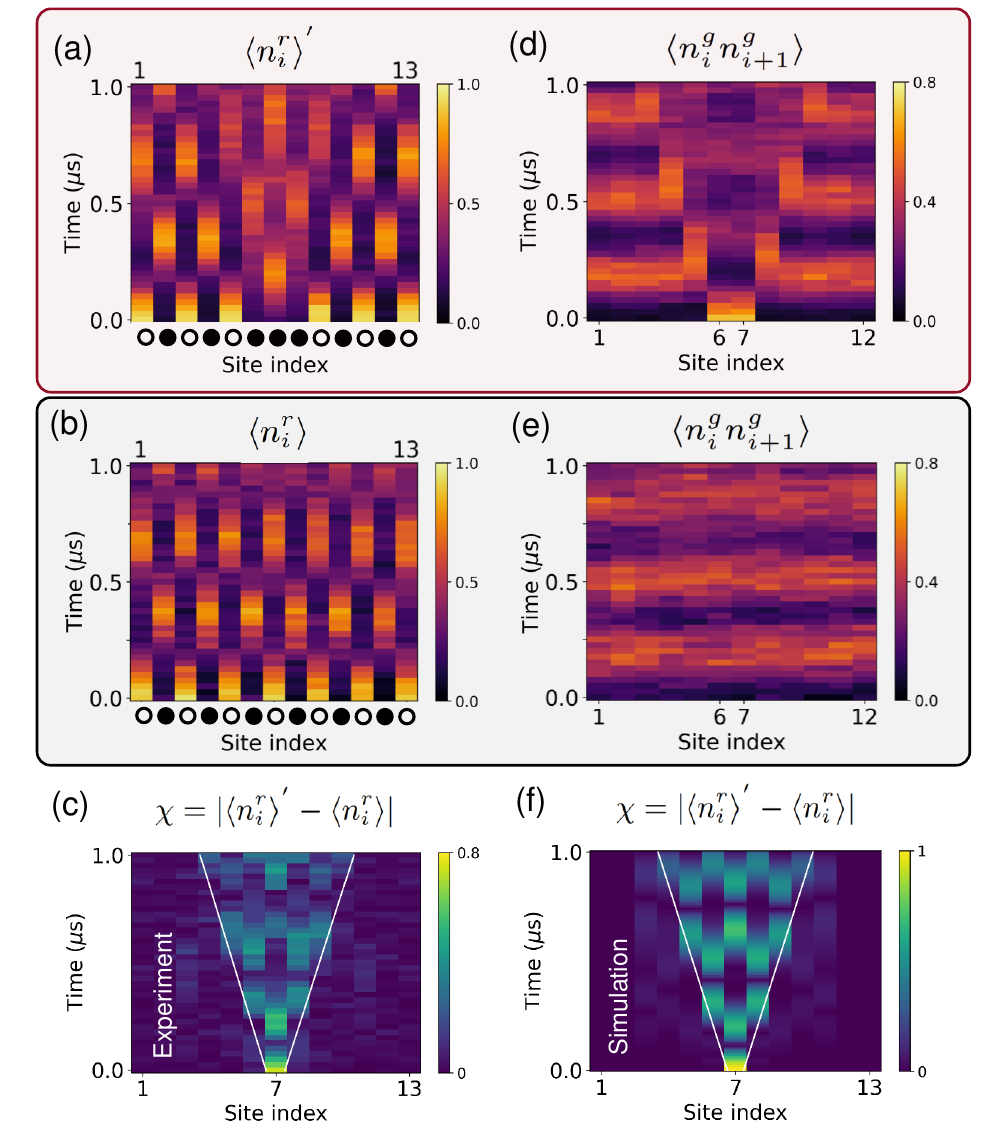}
\caption{\label{fig:holevo}Two experiments are performed to investigate the dynamics of information scrambling through local density and correlation propagation, employing initial states $| {\bf Z}_2\rangle$ and $| {\bf Z}_2^{'}\rangle= \sigma_x^{(\text{7})}| {\bf Z}_2\rangle$, which differs by a local spin flip to the central atom. Panels (a) and (b) present a comparative analysis of experimentally measured local Rydberg atom density, highlighting the nonlocal differences induced from the local perturbation in the initial state following the dynamical evolution. Panels (c) and (f) further compare the local Rydberg atom density differences-experimental data versus numerical simulations. The analogous honeycomblike synchronized structure within the light cone mirrors the characteristic pattern observed for the ZZ-OTOC.
Panels (d) and (e) depict the NN correlator of the ground state density, $ \langle n_i^g n_{i+1}^g \rangle$,  which is interpreted as an antiferromagnetic domain wall density, from the same experiments as in (a) and (b).
}
\end{figure}

Figures \ref{fig:holevo}(a) and \ref{fig:holevo}(b) compare the dynamics of the initial $| {\bf Z}_2^{'}\rangle= \sigma_x^{(\text{7})}| {\bf Z}_2\rangle$ and $|{\bf Z}_2\rangle$ states, respectively. The former is prepared by flipping the Rydberg atom at the central site to its ground state. We find this local disturbance in the initial state disperses into the entire system and the local Rydberg atom density presented in Fig. \ref{fig:holevo}(a) 
reveals similar persistent oscillations and distinct light cone boundaries. Both inside and outside of these boundaries, the oscillations exhibit varying rhythms and amplitudes. 

From the original data in Figs. \ref{fig:holevo}(a) and \ref{fig:holevo}(b), we obtain Fig. \ref{fig:holevo}(c) by calculating the difference of the local Rydberg atom densities between the two experiments
\begin{equation}
    \chi=\lvert\langle n_i^r\rangle^{'} - \langle n_i^r\rangle\rvert, 
    \label{eq:holevo}
\end{equation}
where $\langle n_i^r\rangle^{'}$ and $\langle n_i^r\rangle$ are the expectation values of local Rydberg atom density operator of the initial states $|{\bf Z}_2^{'}\rangle $ and $| {\bf Z}_2\rangle$ respectively. 
These results agree well with the numerical results in Fig. \ref{fig:holevo}(f) and are shown to capture the same essential features of anomalous information scrambling as for the ZZ-OTOC, including the slower information spreading and the persistent oscillations.

We further explore the transport of two-body correlations~\cite{Jurcevic2014Quasiparticle, Richerme2014Non-local} as indicators of information scrambling. Figures \ref{fig:holevo}(d) and \ref{fig:holevo}(e) show the correlations between NNs $\langle n_i^gn_{i+1}^g\rangle$, where $n_i^g=(1-\sigma_{z}^{(i)})/{2}$ is the local ground state atom density at site $i$. Given the constraint imposed by Rydberg blockade,
one expects $ \langle n_i^rn_{i+1}^r\rangle=0$, and the NN correlator hence reflects a ferromagnetic correlation against the background of a $Z_2$ Néel state:
    $\langle n_i^gn_{i+1}^g\rangle=\frac{1}{2}(1+\langle \sigma_{z}^{(i)}\sigma_{z}^{(i+1)}\rangle)$.
Consequently, this data 
can be interpreted as the propagation of domain walls acting as quasiparticle excitations in an antiferromagnetic system. Specifically, two domain walls are initially created at the center by the spin flip and subsequently observed to propagate out toward the boundaries. The propagation speed for the light cone boundaries is 
$1.6 \pm 0.1$, normalized by the Rabi frequency, which is 
comparable to the value 
$1.4 \pm 0.3$
observed for the ZZ-OTOC
. Notably, this speed is 
much lower than the
butterfly velocity {$4.7 \pm 0.2$} in the ZZ-OTOC experiment
for the typical state $\left|\bold{g}\right>$, indicating a slower rate of information spreading. 

\paragraph{Discussions—}

Finally, we remark that, due to the time-forward and -backward Hamiltonian evolution and interleaved local operator, the persistent oscillations of the ZZ-OTOC inside the light cone in general cannot be directly deduced from the eigenstate decomposition of initial states. As a result, our observed oscillations of the ZZ-OTOC are distinct from previously reported periodic revivals of local densities~\cite{Bernien2017Probing, Bluvstein2021Controlling, Zhang2022Many, Su2023Observation} in essential ways. They provide valuable insights into the intricate dynamics of quantum many-body scars and their pivotal roles in information scrambling.
The unique OTOC signatures of $Z_2$ Néel states~\cite{Yuan2022Quantum}, namely, the slower butterfly velocity and the persistent
oscillations within the light cone are observed for the first time. The latter feature is in stark contrast to the rapid decay of the oscillation amplitude inside the light cone observed for randomly selected states from the Hilbert space in this model as well as in other chaotic systems. In contrast to full ergodicity-breaking systems, such as in many-body localization with a logarithmic light cone, information spreading in the Rydberg atom array maintains a linear expansion.
These signatures are imprinted with the kinetically constrained Hamiltonian and further corroborate with measurements of the local Rydberg atom density difference and the spreading of two-body correlations. 

In this work, the challenge for time-reversed evolution is bypassed by 
taking advantage of the inherent constraint imposed by the strong van der Waals interaction. Such a technique is potentially extendable to the nonresonant regime, where synthetic spin exchange interactions emerge from blockade constraints in second-order perturbation theory~\cite{Yang2019Quantum}. 
Our demonstrated techniques in measuring ZZ-OTOC for the Rydberg atom array platform would prove valuable to a variety of other studies including Hilbert space fragmentation~\cite{moudgalya2022quantum,Bucca2022OTOC}, quantum memory~\cite{Bao2024Creating}, and quantum sensing~\cite{Macr2016Loschmidt}. 

\paragraph{Note added—}Recently, we became aware of related work by the group of Prof. Lin Li at HUST, which explores similar quantum information scrambling phenomena.

\begin{acknowledgments}
\paragraph{Acknowledgments—}
We acknowledge significant help and enlightening discussions with Dong-Ling Deng and Dong Yuan. We acknowledge helpful contributions by Xiaoling Wu, Songtao Huang, Yuanjiang Tang, Chenyuan Li and Xiangliang Li in the early stages of building the experimental platform. X. L. also acknowledges constructive discussions with Fan Yang and Guoxian Su.
This work is supported by the Innovation Program for Quantum Science and Technology (2021ZD0302100).
L. Y. is also supported by NSFC (Grants No. 12361131576 and No. 92265205). M. K. T. is supported by NFSC (Grants. No.12234012 and No.W2431002).

\end{acknowledgments}

\bibliographystyle{apsrev4-1}
\bibliography{main}

\end{document}


\preprint{APS/123-QED}

\title{Supplementary Material: Observation of Anomalous Information Scrambling in a Rydberg Atom Array}

\author{Xinhui Liang}
\thanks{These authors contributed equally to this work.}
\affiliation{State Key Laboratory of Low Dimensional Quantum Physics, Department of Physics, Tsinghua University, Beijing 100084, China}
\affiliation{Beijing Academy of Quantum Information Sciences, Beijing 100193, China}

\author{Zongpei Yue}
\thanks{These authors contributed equally to this work.}
\affiliation{State Key Laboratory of Low Dimensional Quantum Physics, Department of Physics, Tsinghua University, Beijing 100084, China}
\affiliation{Frontier Science Center for Quantum Information, Beijing 100084, China}

\author{Yu-Xin Chao}
\affiliation{State Key Laboratory of Low Dimensional Quantum Physics, Department of Physics, Tsinghua University, Beijing 100084, China}

\author{Zhen-Xing Hua}
\affiliation{State Key Laboratory of Low Dimensional Quantum Physics, Department of Physics, Tsinghua University, Beijing 100084, China}

\author{Yige Lin}
\affiliation{Division of Time and Frequency Metrology, National Institute of Metrology, Beijing 100029, China}

\author{Meng Khoon Tey}
\email{mengkhoon_tey@mail.tsinghua.edu.cn}
\affiliation{State Key Laboratory of Low Dimensional Quantum Physics, Department of Physics, Tsinghua University, Beijing 100084, China}
\affiliation{Frontier Science Center for Quantum Information, Beijing 100084, China}
\affiliation{Hefei National Laboratory, Hefei, Anhui 230088, China}

\author{Li You}
\email{lyou@mail.tsinghua.edu.cn}
\affiliation{State Key Laboratory of Low Dimensional Quantum Physics, Department of Physics, Tsinghua University, Beijing 100084, China}
\affiliation{Beijing Academy of Quantum Information Sciences, Beijing 100193, China}
\affiliation{Frontier Science Center for Quantum Information, Beijing 100084, China}
\affiliation{Hefei National Laboratory, Hefei, Anhui 230088, China}

\date{July 31, 2025}

\maketitle

\tableofcontents

\section{Introduction to OTOC} \label{app:otoc}

Quantum information scrambling is commonly measured by the four-point correlator with an unusual time ordering, defined as
\begin{eqnarray}
F(t)=\langle\psi|W^\dagger(t)V^\dagger W(t)V|\psi\rangle,
\label{eq:def}
\end{eqnarray}
where $W(t) = e^{i\mathcal{H}t}We^{-i\mathcal{H}t}$ represents the operator's evolution in the Heisenberg picture.
This expression measures the overlap between the two states $W(t)V|\psi\rangle$ and $VW(t)|\psi\rangle$, which share the same initial state but differ in the sequence of operations applied.
For unitary operators, the above OTOC connects to the squared commutator
\begin{eqnarray}
C(t) = \langle\psi|[W(t),V]^{\dagger}[W(t),V]|\psi\rangle,
\label{eq:def2}
\end{eqnarray}
 through the relation $C(t)=2\{1-\mathrm{Re}[F(t)]\}$, emphasizing the development of noncommutativity between two quantum operations over time.

 The spatial propagation of local quantum operators can be monitored by selecting $W$ and $V$ as local operators, exemplified by  Pauli matrices $\sigma_{\mu}^{(i)}, (\mu=x,y,z)$ of the $i$-th qubit. The commutator $[{\sigma}_\mu^{(i)}(t),{\sigma}_\nu^{(j)}]$ in the space-time diagram illustrates how information transports from one local site to another. In a system with local interaction, the emergence of a light cone with a finite velocity affirms the so-called Lieb-Robinson bounds for information propagation. 
 
 More generally, the spreading of local information does not always adhere to a ballistic pattern but can exhibit diverse growth behaviors, thereby can be employed to characterize different quantum dynamical phases~\cite{Xu2024Scrambling}. As the long-range connectivity of a system increases, the shape of the information light cone transforms from ballistic to exponential form. The corresponding OTOC exhibits rapidly decaying oscillation amplitude inside the light cone and eventually converges to a steady value, as a signature of many-body quantum chaos. Conversely, increasing disorder in a system leads to a slowdown in information propagation, resulting in a logarithmic shaped light cone in a many-body localized system. However, the structure inside the lightcore is rarely studied and it was predicted theoretically that persistent oscillations can occur starting from $Z_2$ Néel state~\cite{Yuan2022Quantum}.

OTOC in general cannot be directly observed in an analog quantum simulator. Under some conditions, however, for instance when the initial state is an eigenstate of the operator $V$(with eigenvalue $v$),  the expression for the OTOC in Eq.~(\ref{eq:def}) simplifies into
\begin{eqnarray}
F(t)=v\langle\Psi(t)|V^{\dagger}|\Psi(t)\rangle
\label{eq:simplify},
\end{eqnarray}
where $|\Psi(t)\rangle = e^{-i\mathcal{H}_{-}t}We^{-i\mathcal{H}_{+}t}|\psi\rangle $, with $H_+$ the time-forward evolution Hamiltonian, and $H_-=-{H}_{+}$ for the time-reversed evolution. This opens up the possible measurement of OTOC experimentally through the properties of final state in a Loschmidt echo experiment~\cite{Quan2006Decay, Macr2016Loschmidt, Jochen22Probing}.

In the OTOC measurement reported in this work, the operators $W$ and $V$ are chosen as local Pauli operators $\sigma_{\mu}^{(i)}$ and $\sigma_{\nu}^{(j)}$ (where $\mu,\nu \in \{x,y,z\}$) at lattice sites $i$ and $j$, respectively, to provide additional spatial resolution for the OTOC $F_{ij}^{\mu\nu}(t)$. 
We choose to measure ZZ-OTOC ($W = \sigma_z^{(i)}, V = \sigma_z^{(j)}$). The operation $\sigma_z^{(i)}$ is carried out by using ac-Stark shift from a phase imprinting laser addressing the atom at site $i$ and $\sigma_z^{(j)}$ corresponds measurement carried out in the Z base at site $j$. The corresponding OTOC then takes the following simple form
\begin{subequations}
\label{eq:whole}
\begin{align}
F_{ij}^{ZZ}(t)
&=(-1)^{\langle n_j^r(0)\rangle+1}[2\langle n_j^r(t)\rangle_i-1],
\label{subeq:1}
\end{align}
\begin{align}
C_{ij}^{ZZ}(t)
&=| \langle n_j^r(t)\rangle_{i} - \langle n_j^r(0)\rangle |
,
\end{align}
\end{subequations}
where $\langle n_j^r(0)\rangle = {\langle\psi_0|n_j^r|\psi_0\rangle}$ and $\langle n_j^r(t)\rangle_{i} = \langle\Psi_{i}(t)|n_j^r|\Psi_{i}(t)\rangle$ are the Rydberg densities of the initial states and the evolved states at site $j$ after total echo time $2t$ following phase imprinting operation at site $i$ respectively.

\section{OTOC for special initial states} \label{app:OTOCspetial}

\begin{figure}[!htp]
\centering
\includegraphics[width=\linewidth]{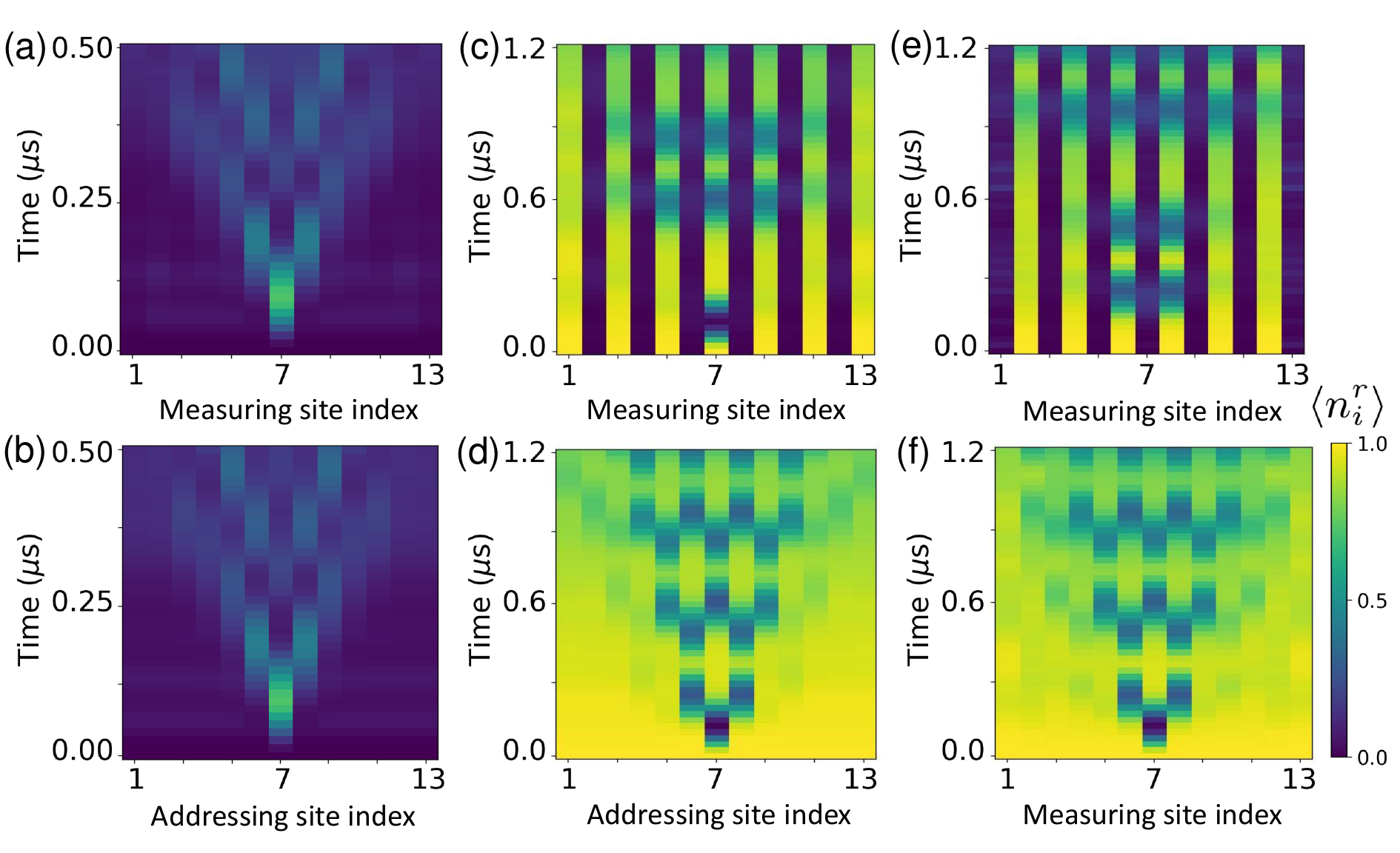}
\caption{\label{fig:otoc_combine} 
Comparisons of the different OTOCs to be measured in our experiment. These results of the Rydberg state density (see the color bar) come from numerically simulated evolutions of the actual Rydberg Hamiltonian (in the following Eq. (\ref{eq:rydberg})) for a 13-atom chain over time with different initial states.
Panels (a) and (b) depict results for the initial states  $\left|\bold{g}\right>=\left|gg...g\right>$, while panels (c) through (f) show results for the ${\bf Z}_2$ Néel states. In (a) and (c) the measurement site index $j$ is changed, while in (b) and (d) the addressing site index $i$ is scanned. Panel (e) mirrors the procedure of (c) but employs the complementary ${\bf \bar Z}_2$ state $\left|grgr...grg\right> $ in place of the ${\bf Z}_2$ state $\left|rgrg...rgr\right>$ in (c). Panel (f) combines data from (c) and (e). Based on these results, it becomes evident that panels (a) and (b) as well as (d) and (f) exhibit similar structures in information scrambling. Minor discrepancies in numerical values are attributed to finite size effects and are negligible during the early stages of evolution, before information spreading reaches the open boundaries.
}
\end{figure}


Specifically for the ZZ-OTOC, there are two direct approaches to visualize information spreading with OTOC, the dependence on the addressing site index $i$ or on the measuring site index $j$, when we focus on the non-commutativity of the operators $F_{ij}$. Notably, for the state $\left|\bold{g}\right>=\left|gg...g\right>$, the two OTOCs exhibit similar structures. In our experiment, we choose to vary the measuring site index $j$ to leverage on the highly paralleled measurements in the Rydberg atom array.

For Néel state, the OTOC measurement calls for slight modification. We find that varying the addressing site index $i$ provides better contrast, as was discussed before \cite{Yuan2022Quantum}. To fully utilize the parallel measurement capabilities, we prepare two complementary Néel states $|{\bf Z}_2\rangle$ and $|{\bf {\bar Z}}_2\rangle$ to obtain the OTOC for odd- and even-numbered sites, respectively. Their measurements are then combined to form an analogous structure similar to that of varying the addressing site $i$ as shown in Fig. \ref{fig:otoc_combine}. 

\section{Experimental setup}

\subsection{Atom rearrangement}


To create a defect-free 1D atomic array of length \( N \), we generate a \( 2 \times N \) optical tweezer array using a spatial light modulator (SLM) and all individual tweezers are
randomly loaded from an overlapping magneto-optical trap (MOT). A defect-free array is achieved through an imaging and rearrangement process, with atom relocation performed using a movable optical tweezer generated by an independent acousto-optic deflector (AOD). The average single-atom rearrangement success rate is $98\%$, resulting in an overall success rate $>70\%$ for the target 15-atom array. For each data point, we perform 1000-2000 experimental runs, post-select successfully rearranged data, and analyze the resulting 700-1500 valid snapshots. Most measurements involve local operators, such as the Rydberg state density \( n_i^r \), for which we simply compute the averages, standard deviations, and error bars according to statistics. The error for a measured operator \( O \) is estimated as $
   \text{Error}(O) = \frac{\sigma_O}{\sqrt{M}}
   $, 
   where \( \sigma_O \) is the standard deviation of \( O \) and \( M \) is the number of valid snapshots. For composite quantities like ZZ-OTOC, defined as the difference in Rydberg state density before and after evolution, we compute errors using error propagation formulas. 

\subsection{Experimental time sequence} \label{app:sequence}

\begin{figure}[h]
\centering
\includegraphics[width=\linewidth]{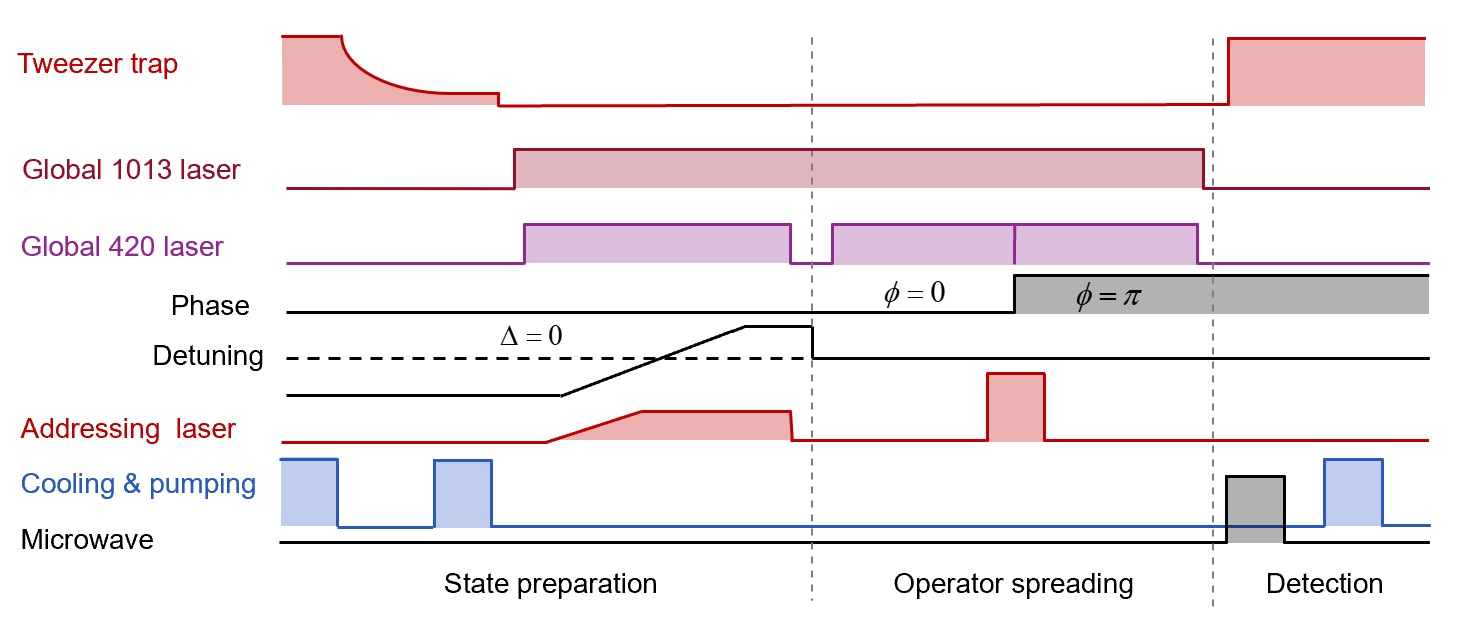}
\caption{\label{fig:sequence} Time sequence of the experiment.
}
\end{figure}

The experimental time sequence for measuring OTOC consists of three main 
 steps as in generic Loschmidt echo or interferometry: preparation of the initial state, analog-digital hybrid evolution to spread the encoded operator, and single-atom resolved final state detection (Fig. \ref{fig:sequence}). All the operations are carried out while the tweezers are turned off and atoms in free fly. 

\begin{figure}[h]
\centering
\includegraphics[width=\linewidth]{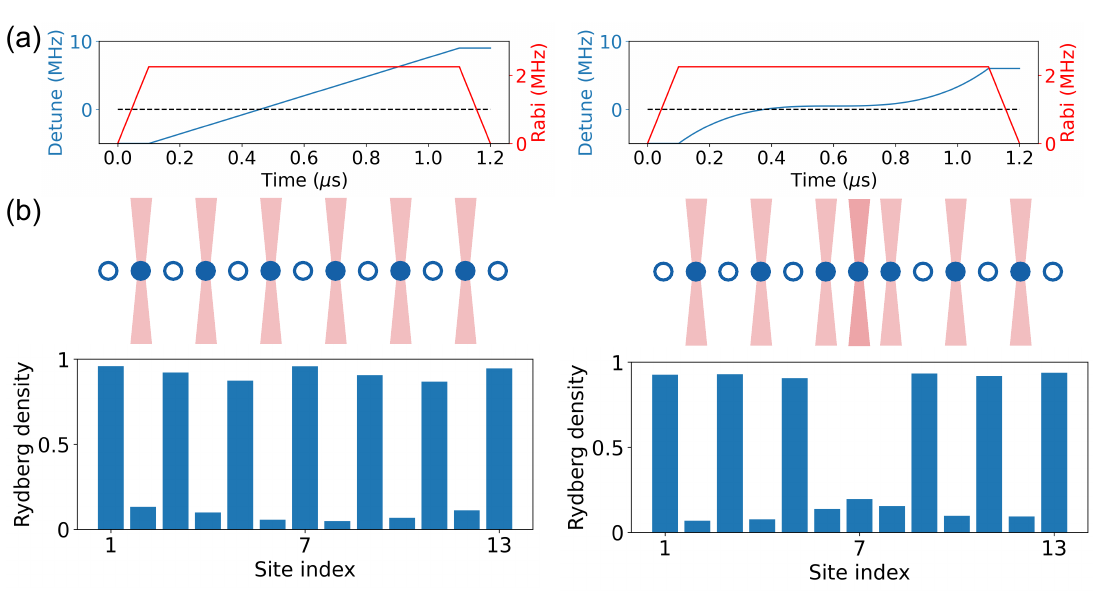}
\caption{\label{fig:Initial state preparetion} Initial state preparation of the $|{\bf Z}_2\rangle$ state and $| {\bf Z}_2^{'}\rangle= \sigma_x^{(\text{7})}| {\bf Z}_2\rangle$ state with one fewer Rydberg atom at the central site. (a) The $|{\bf Z}_2\rangle$ state is prepared by adiabatic sweeping of laser parameters
. (b) The $| {\bf Z}_2^{'}\rangle$ state is prepared as in (a) but the addressing laser pins the central spin to the ground state throughout the adiabatic ramp. 
}
\end{figure}

To prepare the initial state, all atoms are laser cooled and optically pumped into the desired ground state. To prepare the $Z_2$ antiferromagnetic Néel state, the two-photon detuning of the 420-nm and 1013-nm lasers is adiabatically swept from large negative to large positive values. The imperfections can come from the non-adiabaticity of the evolution process as well as the high-frequency noise in the system. Due to the small energy gap in the many-body system, the adiabatic evolution further imposes stringent requirements. Better preparation fidelity is achieved by introducing addressing lasers to pin down the selected atoms, forcing them to stay in the ground state, with every single atom undergoing an analogous Landau-Zener process. The addressing strategy, adiabatic laser frequency sweep, and experimental results of state preparation are presented in Fig.~\ref{fig:Initial state preparetion}.

The global 1013-nm and 420-nm lasers are turned on during the time-forward evolution, and the time-reversed evolution is realized by a sudden $\pi$ phase jump to the 420-nm laser, 
which introduces a negative sign to the effective PXP Hamiltonian. The phase jump is directly programmed into the arbitrary waveform generator (AWG), which controls the phase of the 420-nm laser through a single-pass acousto-optic modulator (AOM). A brief dip of around 10 ns appears in the laser power due to the drop in diffraction efficiency of the AOM during the sudden phase shift. Nevertheless, this time duration is so short that the undesired effect can be neglected over the intended evolution time. Phase imprinting is achieved using the same addressing lasers for preparation of the $Z_2$ initial state.

Single-atom resolved state detection is carried out by switching on the trapping tweezers. Ground state atoms are recaptured, while Rydberg atoms are lost. The recaptured atoms are detected by fluorescence imaging. We use a strong microwave pulse to ionize the Rydberg states in a short time before recapture to enhance detection efficiency.

\begin{figure}[h]
\centering
\includegraphics[width=0.5\linewidth]{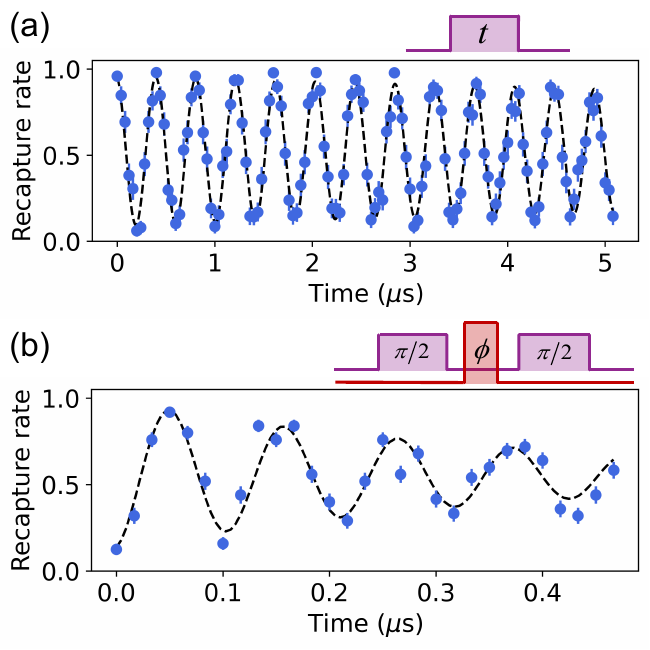}
\caption{\label{fig:phase} Calibration for single qubit operation. (a) Global Rabi oscillation. (b) Addressed site phase imprinting $\phi = \Delta_i^{\text{ac}} t$.
}
\end{figure}

\subsection{Calibration of single-qubit Rabi oscillation and phase imprinting}\label{app:phase}

The single-qubit Rabi oscillation is driven by counter-propagating global 420-nm and 1013-nm lasers with effective Rabi frequency from 2-2.5 MHz as shown in Fig.~\ref{fig:phase}(a).

The single-qubit Z operation is realized via an ac-Stark shift induced by an addressing laser. 
The amount of phase imprinted is calibrated with a single-atom Ramsey experiment with two $\pi/2$ pulses (Fig.~\ref{fig:phase}(b)). The oscillation period is approximately 0.11 $\mu s$, implying that a 0.055 $\mu s$ Z gate operation corresponds to a $\pi$ pulse.

The addressing phase imprinting exhibits significantly faster damping compared to the global case, owing to atomic position fluctuations relative to the narrow waist of the addressing laser. Enhanced coherence can be achieved by improved cooling and by expanding the waist of the addressing laser during phase imprinting with a focal-length-detunable lens.

\section{Imperfections in time-reversed evolution}\label{app:reversed}

\subsection{Systematic errors in the Hamiltonian mapping}

To quantitatively assess the impact of imperfections in the time-reversed evolution on the OTOC, we performed numerical simulations of the actual Rydberg Hamiltonian
\begin{equation}
    H=\frac{\Omega}2\sum_{i}\sigma_x^{(i)}-\Delta\sum_i n_i^r+\sum_{i<j}V(\mathbf{r}_{ij}){n_i^r}{n_j^r}
    \label{eq:rydberg},
\end{equation}
under various approximations to the system parameters. 

\begin{figure}[!htp]
\centering
\includegraphics[width=0.6\linewidth]{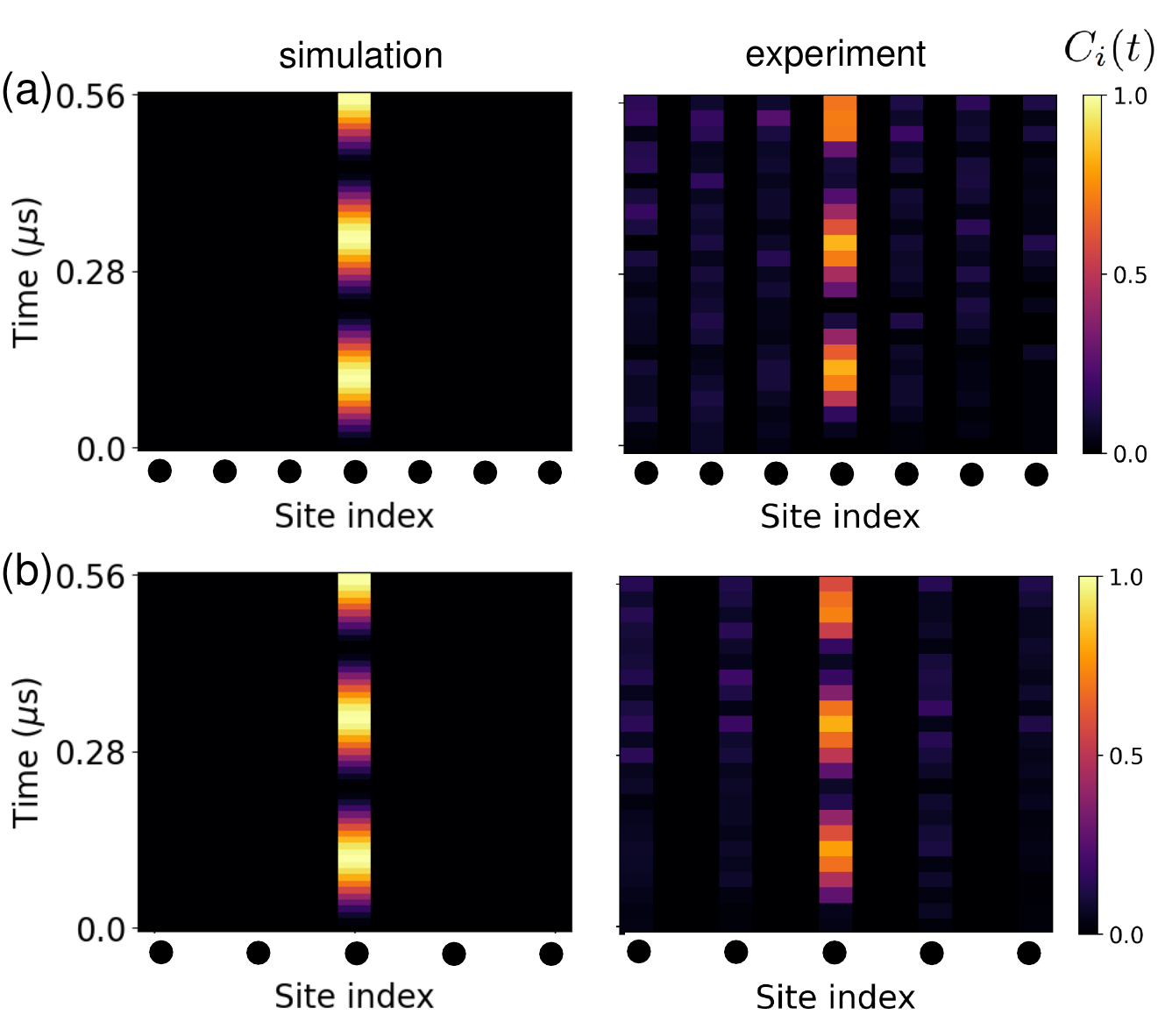}
\caption{\label{fig:single} OTOC of non-interacting atom chain. Compared to the OTOC of the 13 atom array chain from the initial state $\left|\bold{g}\right>=\left|gg...g\right>$ in the main text, the atoms are rearranged to a larger inter-atomic spacing, 
at (a) twice or (b) three times the 6 $\mu m$ spacing used in the main text respectively, significantly diminishing the strength of
NN interactions. The information then concentrates on the addressed atom and cannot spread to other sites without inter-atom interactions.
}
\end{figure}

We first start from non-interecting cases with large inter-atomic spacing. The experimental OTOC results for non-interecting atom array are presented in Fig. \ref{fig:single}. The signal is found to be primarily concentrated on the addressed atom located at the center of the atom array, exhibiting behavior equivalent to a t-$\pi$-t single atom Rabi oscillation. 

We now discuss simulation results based on the actual Rydberg Hamiltonian assuming perfect controls. 
The primary errors due to 
the deviations of the physical Hamiltonian from the ideal PXP model, particularly with respect to the neglect of next-to-nearest-neighbor (NNN) interactions, followed by second-order virtual couplings to states that violate the blockade condition. 

The choice of interatomic spacing in the Rydberg array simulator can be leveraged to balance and minimize the systematic errors in OTOC measurements. Since the ratio between NN and NNN interactions is fixed at 64 to 1, reducing the NNN interactions also reduces the NN blockade strength. Therefore, an optimal atomic spacing exists. 

\begin{figure}[!htp]
\centering
\includegraphics[width=0.7\linewidth]{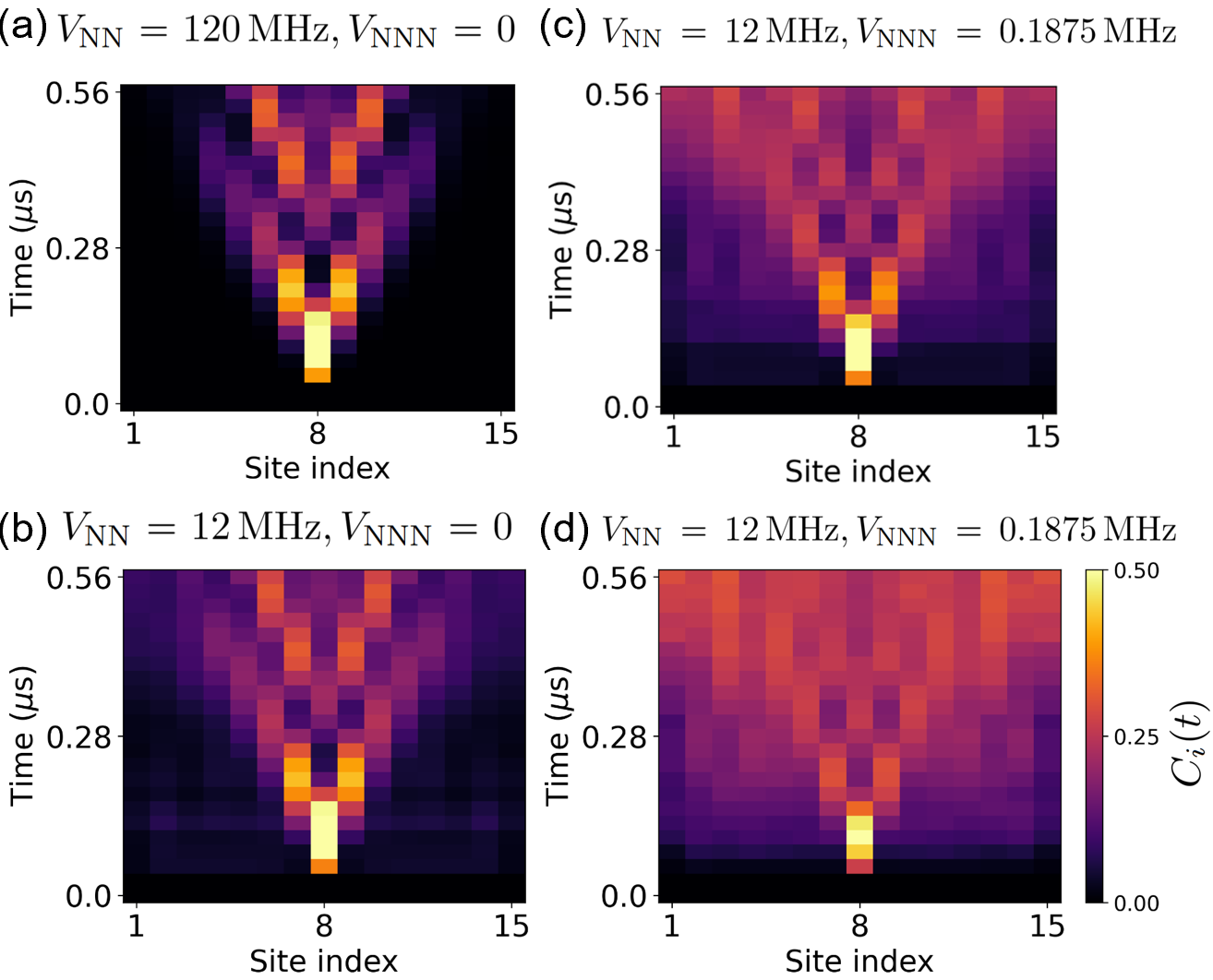}
\caption{\label{fig:error} Numerical results of Rydberg density or $ C_{i}(t)$ for the initial state $\left|\bold{g}\right>=\left|gg...g\right>$ at different parameters $\Delta = 0, \Omega_i= \SI{2.5}{MHz}$. (a) $V_{\rm{NN}}=\SI{120}{MHz}, V_{\rm{NNN}}=0$, (b) $V_{\rm{NN}}=\SI{12}{MHz}, V_{\rm{NNN}}=0$, (c) $V_{\rm{NN}}=\SI{12}{MHz}, V_{\rm{NNN}}=\SI{0.1875}{MHz}$, (d) for the same parameters as in (c) but with noise modeled including variations in atomic position, single-spin dephasing, and depolarization.
}
\end{figure}

We find when the NN interaction strength is sufficiently large at 120 MHz (Fig. \ref{fig:error}a, 10 times the experimental value) and interactions beyond NN are neglected, the system closely mimics the behavior of the PXP Hamiltonian 
\begin{eqnarray}
H=\frac{\Omega}2\sum_{i=1}^{L-1}P_{i-1}\sigma_{x}^{(i)}P_{i+1}
\label{eq:pxp}.
\end{eqnarray} 
In this case, the time-reversed evolution is nearly perfect, resulting in a clear and distinct OTOC signal that easily stands out from the background noise. However, when the blockade condition is relaxed to the NN interaction strength of 12 MHz (Fig. \ref{fig:error}b, the experimental value), we observe a rise in the background noise due to leakage into otherwise forbidden subspaces. Furthermore, the inclusion of NNN interaction (Fig. \ref{fig:error}c), which is 64 times weaker than the NN interaction, introduces even more deviations. Given the fixed ratio between NN and NNN interaction strengths, by carefully selecting a suitable spacing, which in this experiment is set to $6 ~\mu m$, with a resulting NN interaction of 12 MHz, one can optimize the trade-off between NNN interaction and violation of the NN blockade condition, thereby enhancing the accuracy and reliability of the experimental results.



\subsection{Imperfections from technical errors}

\begin{figure}[!htp]
\includegraphics[width=\linewidth]{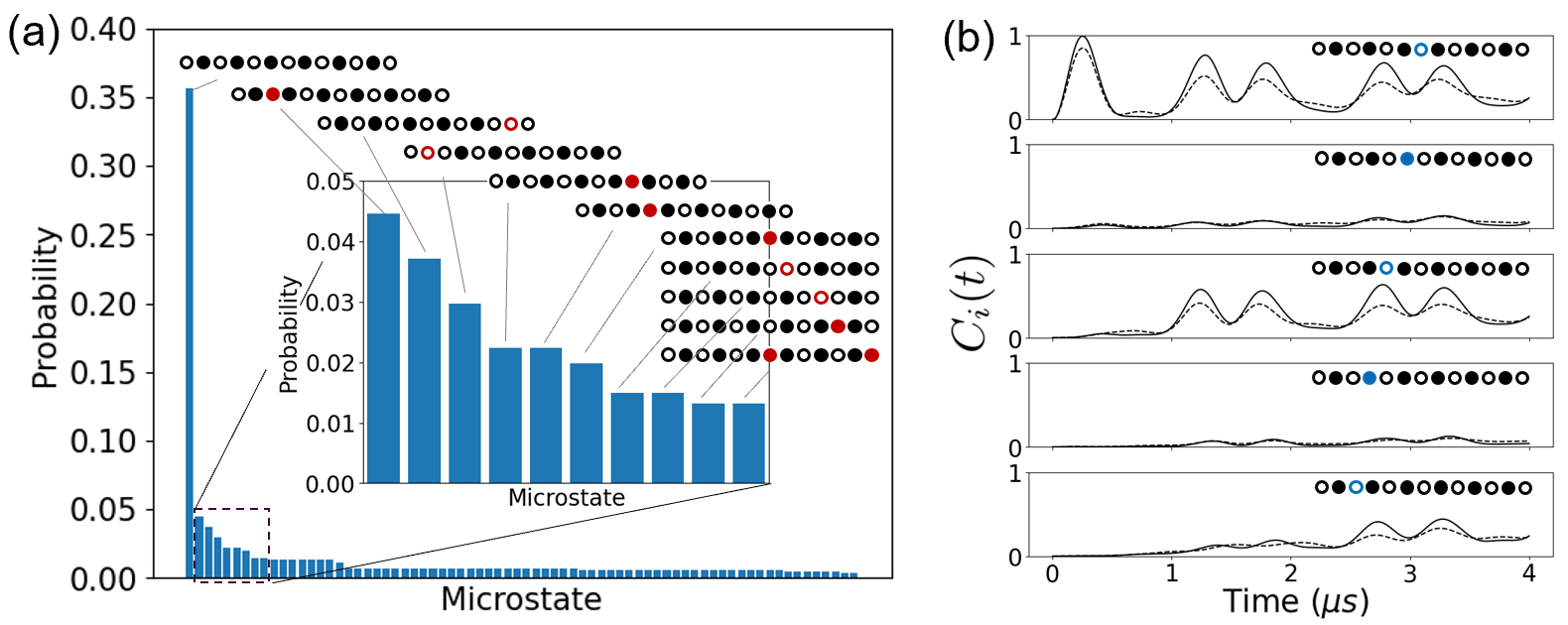}
\caption{\label{fig:stateprepare_error} Estimation of the impact of initial state preparation errors on OTOC results: (a) Probability distribution of the different configurations in the initial state preparation. The atomic spin flips deviate from the intended target $\left|{\bf Z}_2\right\rangle$ Néel state are marked in red. (b) The ideal OTOC (solid line) and the OTOC considering the probability distribution of the initial state preparation (dashed line), estimated numerically by weighted averaging of the OTOCs for each configuration. The blue symbols mark the indexed site spin states for measuring $C_i(t)$.
}
\end{figure}

The observed OTOC contrast is lower than theoretical prediction, after considering the systematic errors in the Hamiltonian mapping. The remaining source of imperfections primarily come from the following experimental technical errors: SPAM errors, atomic position fluctuations which lead to phase injection errors and variations in van der Waals interactions $V(\mathbf{r}_{ij})$, and single-atom decoherence effects. 

Phase injection errors caused by atomic position fluctuations affect the amplitude of the injected signal. Similar to detection errors, they contribute to the contrast reduction of OTOC signals. By measuring and fitting the contrast reduction in OTOC for non-interacting atom arrays (Fig. \ref{fig:single}), we estimate the corresponding contribution to contrast reduction is at about 12\%.

Initial state preparation and imaging loss (i.e. atom loss during imaging) also contribute to reduced contrast. The preparation of the $\left|\bold{g}\right>$ state involves imaging loss and optical pumping infidelity. By fitting the repeated imaging loss curve, we estimate the imaging loss to be less than 1\%. Similarly, the error in a single optical pumping cycle is also less than 1\%, determined by fitting the leakage curve after multiple optical pumping cycles.
These imperfections introduce holes in the state preparation process, which could potentially hinder information propagation. However, the small infidelity 
does not significantly affect the averaged results.

The preparation of the $\left|{\bf Z}_2\right\rangle$ Néel state requires additional steps. Figure ~\ref{fig:stateprepare_error}(a) show the state preparation fidelity and main configurations that differ 
from the target state by a few atomic spin flips. By measuring the probability of each 
configuration in the initial state preparation and analyzing the results for each configuration 
independently, and followed by weighted averaging, we find that initial preparation errors mainly 
affect the OTOC contrast at the corresponding sites in Fig.~\ref{fig:stateprepare_error}(b). 

\subsection{Imperfections from single atom decoherence}\label{sec:decoherence}

The numerical simulation in Fig.~\ref{fig:error}(d) incorporates additional single-atom decoherence mechanisms, which include: (1) depolarization rate derived from longitudinal relaxation time obtained through single-atom fitting, which accounts for Rydberg state lifetime and Rydberg excitation laser scattering, estimated to be approximately \SI{20}{\kilo\hertz}; 
(2) dephasing rate calculated by subtracting the contribution of longitudinal relaxation rete from transverse relaxation rate, yielding a fast dephasing rate of \SI{40}{\kilo\hertz}; and (3) atomic velocity and position uncertainty estimated from temperature measurements. The standard deviation of position fluctuation is approximately 5\% of the interatomic spacing, corresponding to \SI{0.3}{\micro\metre} in \SI{10}{\micro\second} trap release time, while the atomic velocity distribution introduces additional Doppler shift with a standard deviation of $40~\text{kHz}$. 

\begin{figure}[b]
\includegraphics[width=0.8\linewidth]{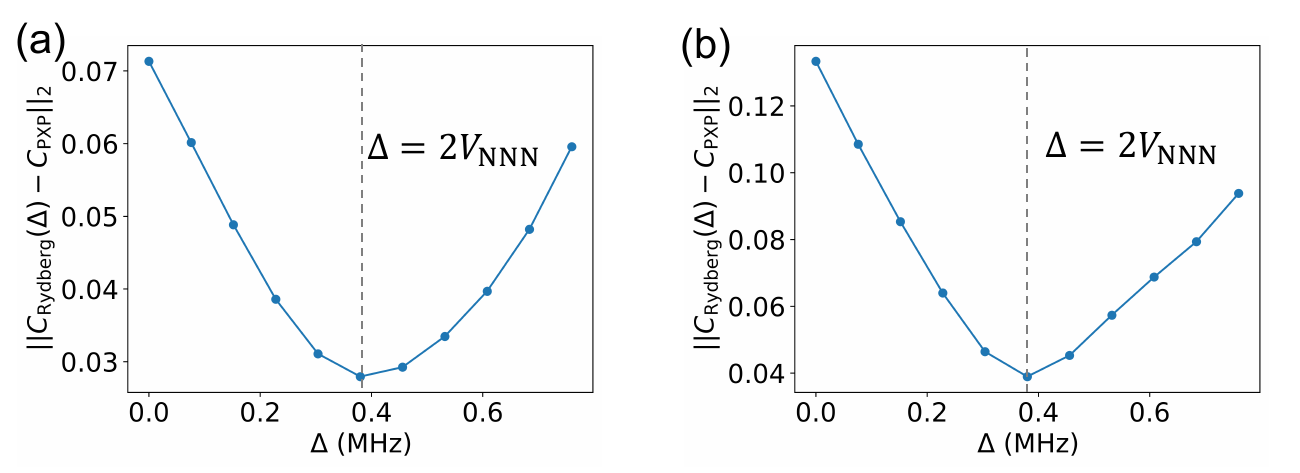}
\caption{\label{fig:compensation} Error compensation. Optimization of the compensation detuning to minimize the distance $|| C_{\rm Rydberg}(\Delta)-C_{\rm PXP}||_2$ between the OTOC calculated with the actual Rydberg array Hamiltonian and that from the ideal PXP Hamiltonian. The optimal detuning is approximately twice the NNN interaction strength, for both (a) $\left|\bold{g}\right>=\left|gg...g\right>$ and (b) $\left|{\bf Z}_2\right>=\left|rgrg...rgr\right>$ studied in the main text.
}
\end{figure}

To account for these variations, we employ Monte Carlo sampling by selecting different position and velocity distributions for atoms in the array for each simulation run, and probabilistically adding depolarization and decoherence noise to the dynamics. The final results are obtained by averaging over 300 samples. Whereas this approach achieves good agreement between simulation results and experimental data for the fully polarized initial state, the experimental results for the $\left|{\bf Z}_2\right\rangle$ Néel state exhibit lower contrast with the same parameters. We attribute this additional degradation to extra state preparation infidelity and atomic heating during the addressed preparation of the initial state, particularly when there is an offset between the addressing light and optical tweezer trapping single atom. This misalignment can significantly accelerate the atom in one direction, resulting in a wider velocity distribution and faster decay rate compared to theoretical predictions.

\section{Error compensation and subtraction}\label{app:deduction}

In our experiments, an extra detuning is introduced to partially cancel out the NNN interactions in the time-reversed evolution as was first suggested in the Ref. \cite{Yuan2022Quantum}. The detuning remains a constant throughout the time-forward and time-reversed evolution and is optimized based on comparisons between numerical results for the ideal PXP model Eq. (\ref{eq:pxp}) and the complete Rydberg Hamiltonian model Eq. (\ref{eq:rydberg}). Notably, for the $\left|\bold{g}\right>$ and both $\left|{\bf Z}_2\right>$ states, we find optimal compensation favors a detuning value that is approximately twice the strength of the NNN interaction (Fig. \ref{fig:compensation}).  It can be understood as using a frequency offset to compensate for the effects of NNN interactions, with the Hamiltonian expressed as
\begin{equation}
H=\frac{\Omega}2\sum_{i}\sigma_x^{(i)}-\sum_i (\Delta-V_{\text{NNN}}({n_{i-2}^r}+{n_{i+2}^r})) n_i^r+\sum_{i,j=i+1}V_{\text{NN}}{n_i^r}{n_j^r}.
\end{equation}
For both the $\left|\bold{g}\right>$ and $\left|{\bf Z}_2\right>$ states, due to their quasi-translational symmetry, $\left|\phi(i)\right>=\left|\phi(i+2)\right>$, and Hamiltonian $H(i)= H(i+2)$, the spins on the NNN sites always remain identical in the evolution driven by global lasers, $\left<n_{i}^r\right>\approx\left<n_{i+2}^r\right>$. For the state with $\left<n_{i}^r\right>=1$, $\left<n_{i-2}^r\right>+\left<n_{i+2}^r\right>\approx2$. By introducing the extra detuning $\Delta=2V_{\text{NNN}}$, the coefficient before the single-site term $n_i^r$ approaches zero ($\Delta-V_{\text{NNN}}({n_{i-2}^r}+{n_{i+2}^r})\approx0$) during both the time-forward and time-reversed evolution, significantly reducing the time-reversal errors caused by NNN interactions.

\begin{figure}[!htp]
\includegraphics[width=\linewidth]{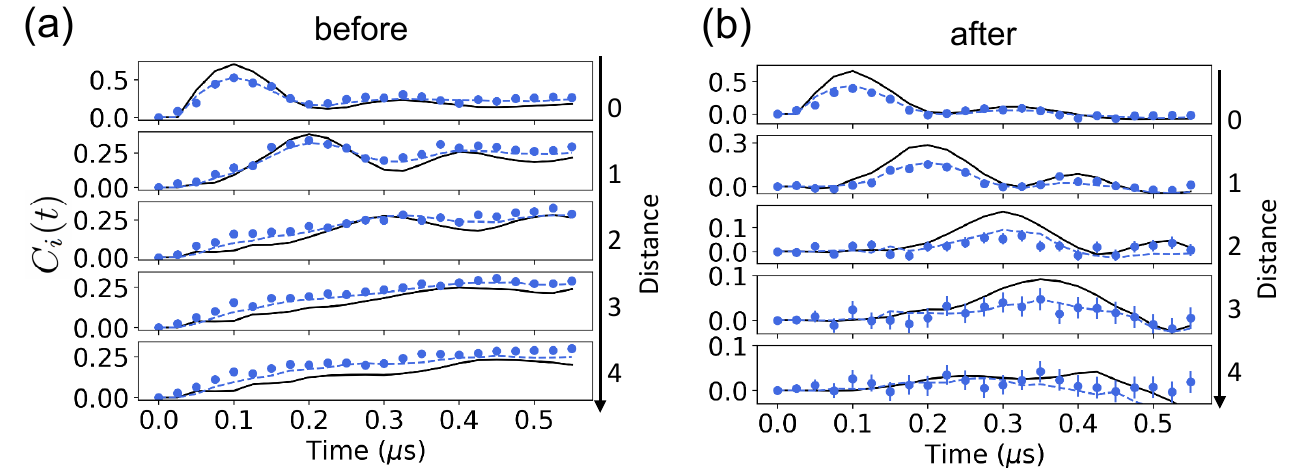}
\caption{\label{fig:deduction} OTOCs (a) before and (b) after error subtraction. Background subtraction is effective in removing the bias introduced by imperfect time-reversal, ensuring that the non-zero values of the OTOC represent genuine information scrambling. Results from experiment (blue dots) are compared with numerical simulation using the actual model Hamiltonian Eq. (\ref{eq:rydberg}) (black solid lines), and with modeled decoherence sources in Sec. \ref{sec:decoherence} (blue 
 dashed lines).
}
\end{figure}

To mitigate other biased errors in the OTOC, the errors stemming from imperfect time-reversed evolution are subtracted. The subtracted background can be quantified through an analogous experiment, but without the central-spin Z operator. In our experiments, atoms situated far away from the central one undergo identical global and time-reversed evolution yet remain unaffected by operator spreading during the evolved time, thus they can be employed as a reference for the background (as in common mode rejection). Compared to repeating the experiment for the sole purpose of obtaining the background error, the approach of referencing to distant atoms further facilitates the elimination of common-mode noise stemming from long-term shifts in experimental parameters. Figure~\ref{fig:deduction} shows the results before and after the discussed error subtraction.

\section{Data analysis and fitting procedures}\label{app:data_analysis}

\subsection{Velocity fitting}

The experimental results for butterfly velocity fitting are based on the procedures illustrated in Fig.~\ref{fig:wavefront_fit}:

\begin{figure}[!htp]
\includegraphics[width=0.8\linewidth]{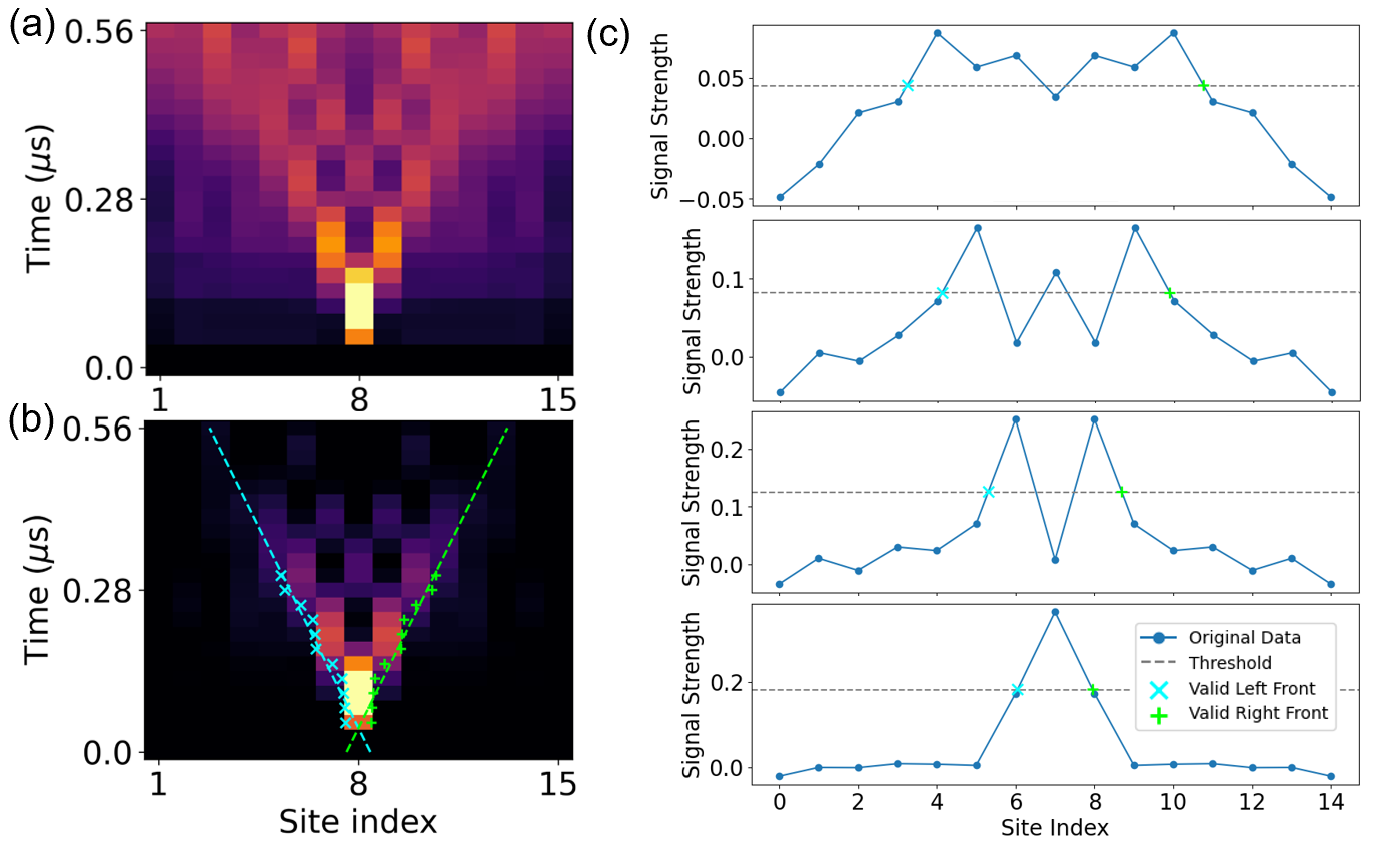}
\caption{\label{fig:wavefront_fit} Velocity fitting procedure. (a) Example of raw data. (b) Processed data after error subtraction with fitted wavefront (solid line). (c) Temporal slice of data from (b) at selected time instants, showing wavefront determination via half-maximum peak position sites.}
\end{figure}

\begin{figure}[!htp]
\includegraphics[width=0.8\linewidth]{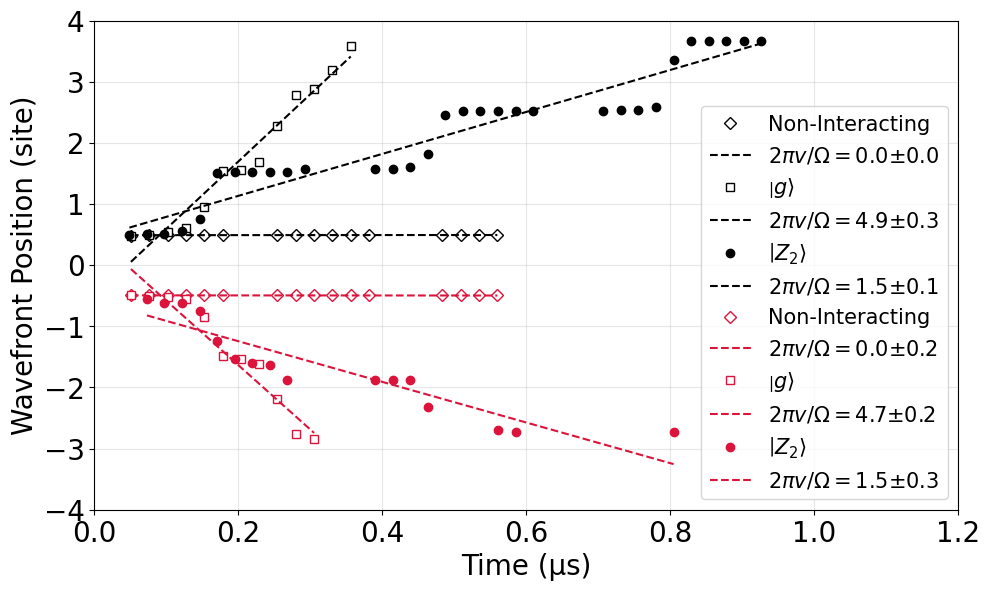}\caption{\label{fig:velocity_fit}
Comparison of OTOC velocity fitting for different initial states and Hamiltonians. The upper half panel shows numerical data in black, while the lower half panel displays experimental data in red as a mirror image for direct comparison.
The dashed lines represent linear fits to the corresponding datasets, with the fitted wavefront propagation velocity $2\pi v/\Omega$ 
indicated in the legend. Our measurements demonstrate distinct velocity characteristics across different quantum states: the non-interacting system (hollow diamonds) exhibits complete absence of signal propagation (v = 0), while comparative analysis indicates the fitted velocity for the $\left|\bold{g}\right>$ initial state (hollow squares) substantially exceeds that of the $\left|{\bf Z}_2\right\rangle$ initial state (solid circles).}
\end{figure}

\begin{itemize}[nosep]
    \item Error subtraction (Sec. \ref{app:deduction}) is first performed to subtract background noise from the data.
    \item AT each time instant, the wavefront is identified by locating the half-maximum position of the peak, as shown in the inset of Fig.~\ref{fig:wavefront_fit}. Data points with no signal (e.g., at zero time) or with signal-to-noise ratio (SNR) too low for reliable fitting (e.g., at long times) are excluded.
    \item All wavefront positions are then linearly fitted to obtain an estimate of the average velocity.
\end{itemize}

\begin{figure}[!htp]
\includegraphics[width=0.7\linewidth]{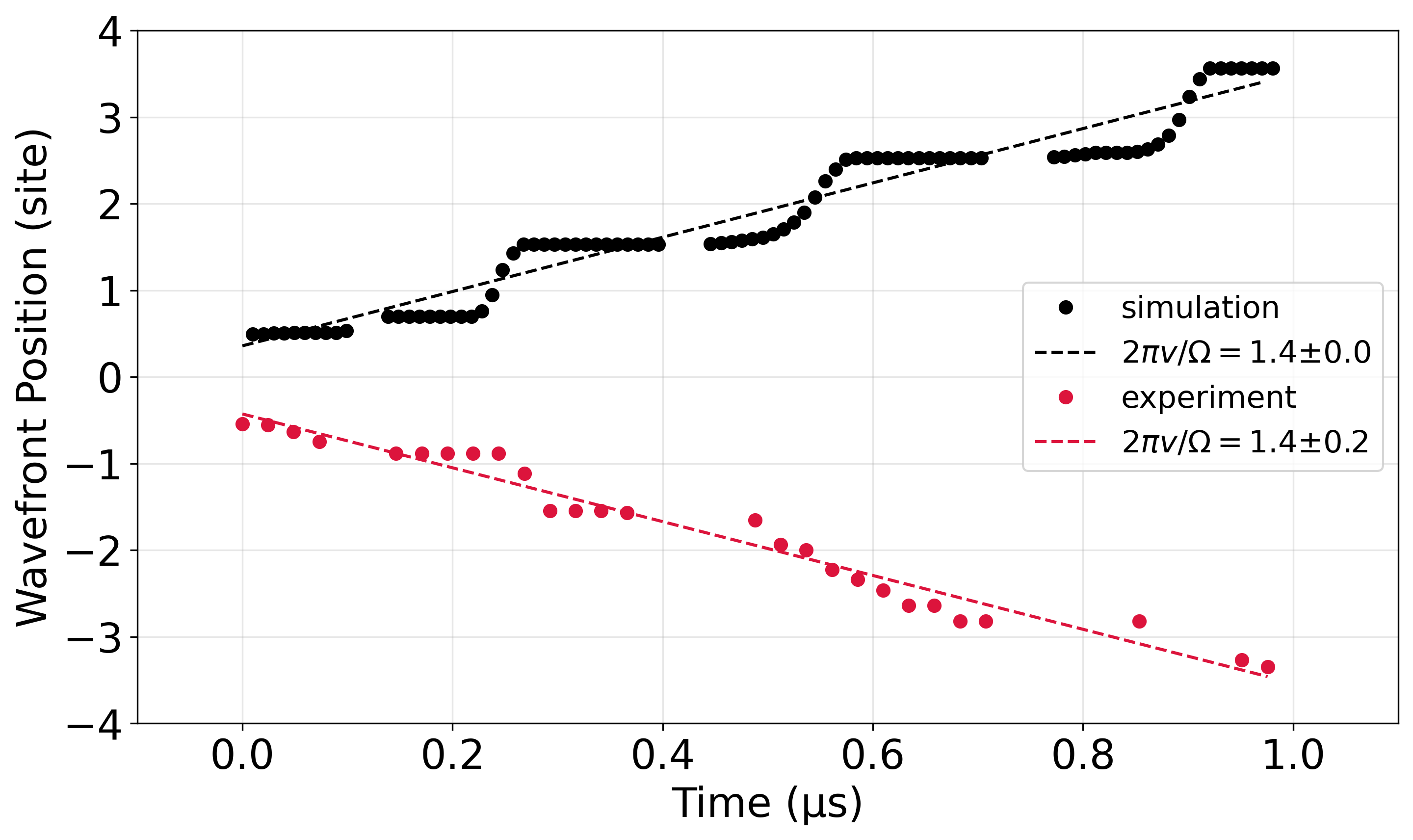}\caption{\label{fig:helovo_velocity}
Velocity analysis for data in Fig. 4 (c,f) of the main text. Upper (lower) half panel depicts numerical (experimental) data. The dashed lines represent linear fits to the corresponding datasets, with the fitted wavefront propagation velocity $2\pi v/\Omega$ normalized by the Rabi frequency indicated in the legend.}
\end{figure}

Figure~\ref{fig:velocity_fit} illustrates a comparison of OTOC velocity fitting results for different initial states and Hamiltonians. The upper panel depicts numerical results, while the lower panel presents the corresponding experimental data mirrored for convenient comparison. Different symbols represent various interaction strengths and initial states, as indicated in the legend: non-interacting system, many-body system with $\left|\bold{g}\right>$ initial state, and many-body system with $\left|{\bf Z}_2\right\rangle$ initial state. Figure \ref{fig:helovo_velocity} demonstrates the same fitting procedure applied to data in Fig. 4 of the main text.

\begin{figure}[!htp]
\includegraphics[width=0.7\linewidth]{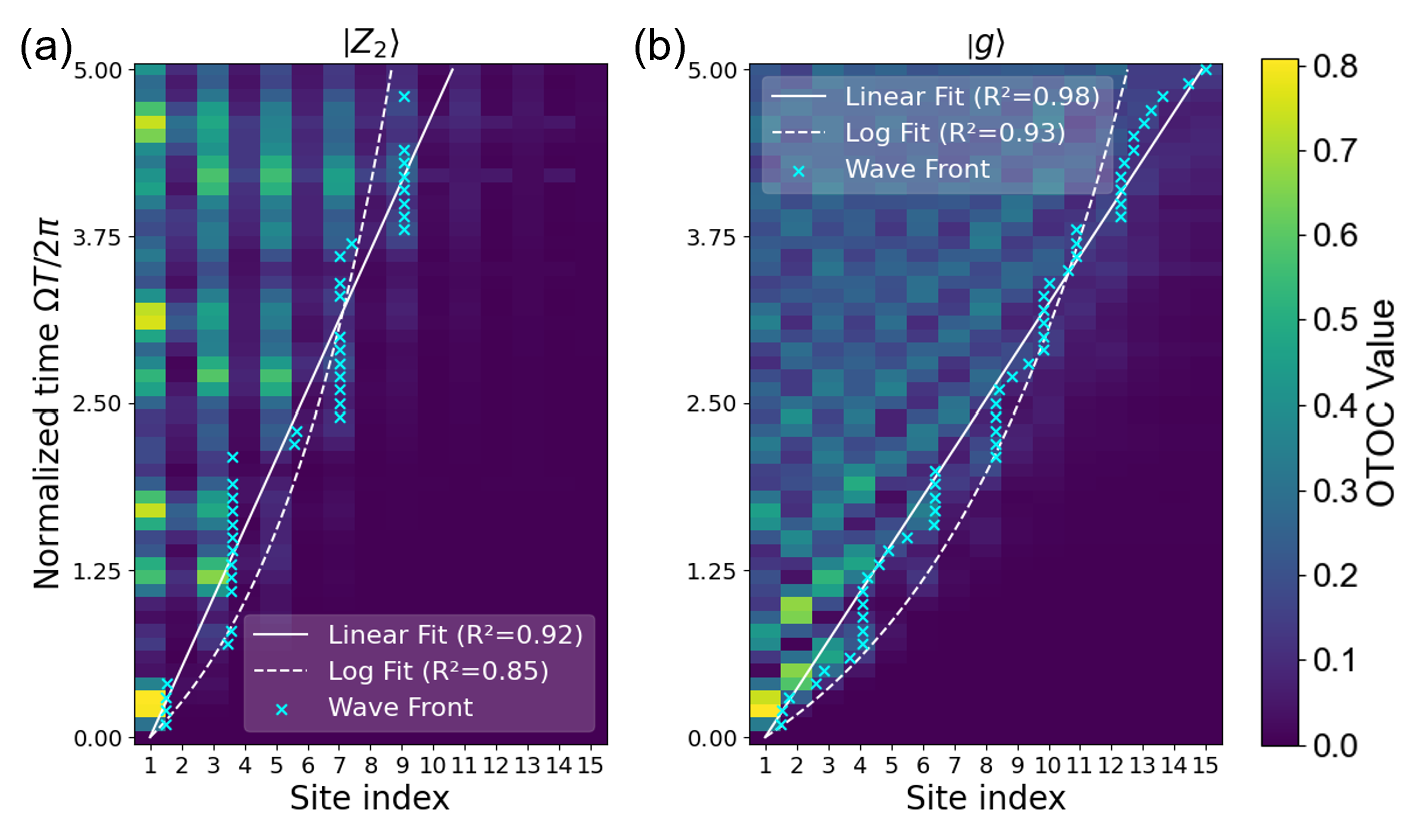}\caption{\label{fig:shape_fit}
Comparison of OTOC wavefront fitting for different line shapes and initial states (a) $\left|{\bf Z}_2\right\rangle$ and (b) $\left|\bold{g}\right>$ based on numerical results. The light blue crosses represent the wavefront to be fitted, while the solid and dashed lines show the optimized fits using linear and logarithmic function forms respectively, each with a single fitting parameter. The excellent agreement of the linear fit compared to the logarithmic alternative quantitatively confirms that the linear function form better captures the wavefront propagation dynamics in our system.}
\end{figure}

The linear fitting is sufficient for capturing short-time wavefront dynamics. We carry out extended numerical simulations for both initial states ($\left|{\bf Z}_2\right\rangle$ and $\left|\bold{g}\right>$) to compare different fitting line shapes (Fig.~\ref{fig:shape_fit}). Using identical one-parameter fitting ($d_{\text{linear}}=vt$, $d_{\text{log}}=v \log{(t+1)}$, with normailized $t=\Omega T/2\pi$), the linear form is found to exceed logarithmic fitting ($R^2_{\text{linear}}=0.92 >  R^2_{\text{log}}=0.85$ for $\left|{\bf Z}_2\right\rangle$ state and 
$R^2_{\text{linear}}=0.98 >  R^2_{\text{log}}=0.93$ for $\left|\bold{g}\right>$ state, where $R^2_{\text{linear/log}}$ denote the coefficient of determination and quantifies the proportion of the variance in the dependent variable that is predictable from the independent variable in a regression model). For the $\left|{\bf Z}_2\right\rangle$ state, better two-parameter fitting ($d=vt^{\alpha}$) with an extra power exponent $\alpha$ of 0.94, a value very close to unity, confirms nearly linear wavefront propagation. The $\left|\bold{g}\right>$ state yields an exponent of 0.86, still very close to linear but exhibiting slight deceleration. 
Subsequent numerical results demonstrate that despite of this deceleration, the velocity remains higher than that of the $\left|{\bf Z}_2\right\rangle$ state.

The deceleration originates from dynamical constraint on Hilbert space induced by Rydberg blockade. For the $\left|{\bf Z}_2\right\rangle$ state, the dynamics immediately encounter restrictions from the extensive Hamming-distance-1 subspace. In contrast, the $\left|\bold{g}\right>$ state initially explores a subspace containing only single Rydberg excitations, unimpeded by blockade effects and thus exhibiting faster information spreading. As evolution progresses to larger Hilbert-space regions, constrained dynamics manifest through gradually decreasing velocities, yet remain consistently exceeding that of the $\left|{\bf Z}_2\right\rangle$ state.

\subsection{Finite-size effects on OTOC and velocity fitting}

\begin{figure}[!htp]
\includegraphics[width=0.9\linewidth]{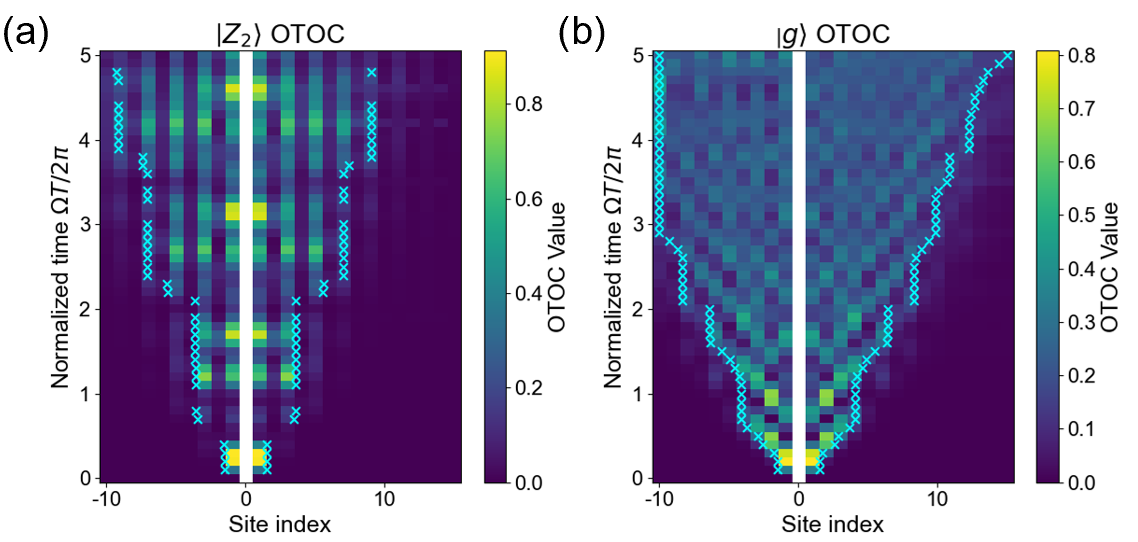}
\caption{\label{fig:finitesize_fit} Comparison of velocity fitting results with different system size and fixed fitting time window. Numerical OTOC results for initial states $\left|{\bf Z}_2\right\rangle$ (a) and $\left|\bold{g}\right>$ (b) for different system sizes (with half the number of total sites $(N+1)/2=$ 10 and 15 sites for left and right sub-panels respectively). Symbols represent fitted wavefront positions at each time instant. The fitted wavefront positions exhibit nearly mirror symmetry between the two different system sizes, demonstrating consistent velocity fitting within the same time window before the wavefront reaches system boundaries. For the $\left|\bold{g}\right>$ initial state, the wavefront reaches the 10-site boundary at normalized time $\Omega T / {2 \pi} = 3$. 
}

\end{figure}

To investigate the influence of finite system size on OTOC results and velocity fitting, we performed numerical simulations for different system sizes (Fig. \ref{fig:finitesize_fit}). By comparing fitting results with same fitting time window ($\Omega T / {2 \pi} \in [0,3]$ for example) for different system sizes, our analysis demonstrates that boundary effects have negligible effect on OTOC dynamics as well as velocity fitting for the time window we choose. Specifically, before information spreading reaches the system boundary, the fitting results from different system sizes are nearly identical. 
Significant differences in fitted velocities only emerge when the information wavefront fully reaches the boundaries of the finite system.

The velocity fitting results depend on the time window used for analysis, which in practice is maximized to minimize fitting errors in the velocity estimation.
In our experiments, we employed the maximum evolution time constrained by the system's coherence and the corresponding system size. While the system size could be larger, increasing the simulation size is found to introduce negligible effects on the results within the current evolution time scale.

\subsection{Decay analysis of OTOC}
\begin{figure}[h]
\includegraphics[width=0.7\linewidth]{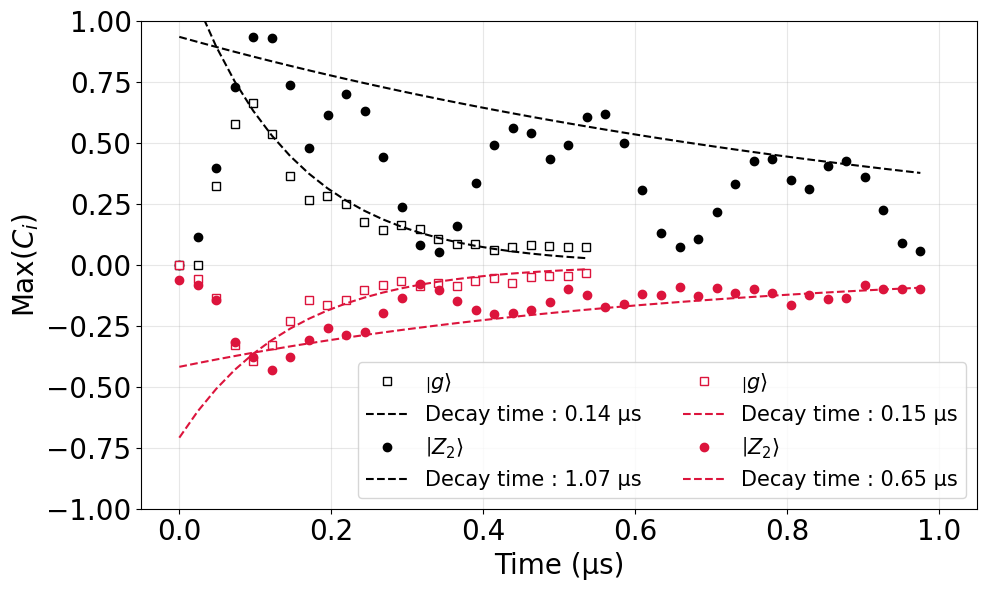}\caption{\label{fig:decay_fit}
Comparison of OTOC decay timescales for different initial states. Upper half panel shows numerical data (black), lower half panel displays experimental data (red) as a mirror image for easier comparison. Dashed lines indicate logarithmic fits to the oscillation envelope with corresponding timescales annotated. The $\left|{\bf Z}_2\right\rangle$ initial state (solid circles) exhibits significantly more pronounced and persistent oscillations compared to the $\left|\bold{g}\right>$ initial state (hollow squares).}
\end{figure}
To quantify the decay timescale of OTOC, we conduct the following fitting procedure:

\begin{itemize}[nosep]
    \item Extract the envelope of OTOC by taking the maximum value $C(t) = \max_i[C_i(t)]$ at each time instant.
    \item Identify the oscillation peaks in the OTOC.
    \item Apply a logarithmic fitting to the extracted peaks to determine the decay timescale.
\end{itemize}

Figure~\ref{fig:decay_fit} shows the fitting of OTOC decay timescales using logarithmic fitting of the oscillation peaks. The results clearly demonstrate distinct decay behaviors for different initial states in the many-body system. Specifically, the $\left|{\bf Z}_2\right\rangle$ initial state exhibits a slower decay rate compared to the $\left|\bold{g}\right>$ initial state, the most dramatic feature being its
oscillatory behavior, implicating different information scrambling dynamics in the quantum states.

\section{Velocity distribution for different initial states}

To demonstrate the representativeness of the two selected initial states and their corresponding propagation velocities in our OTOC experiments, we perform numerical simulations of OTOC results for various initial states.

\subsection{OTOCs for eigenstates}

\begin{figure}[!htp]
\includegraphics[width=\linewidth]{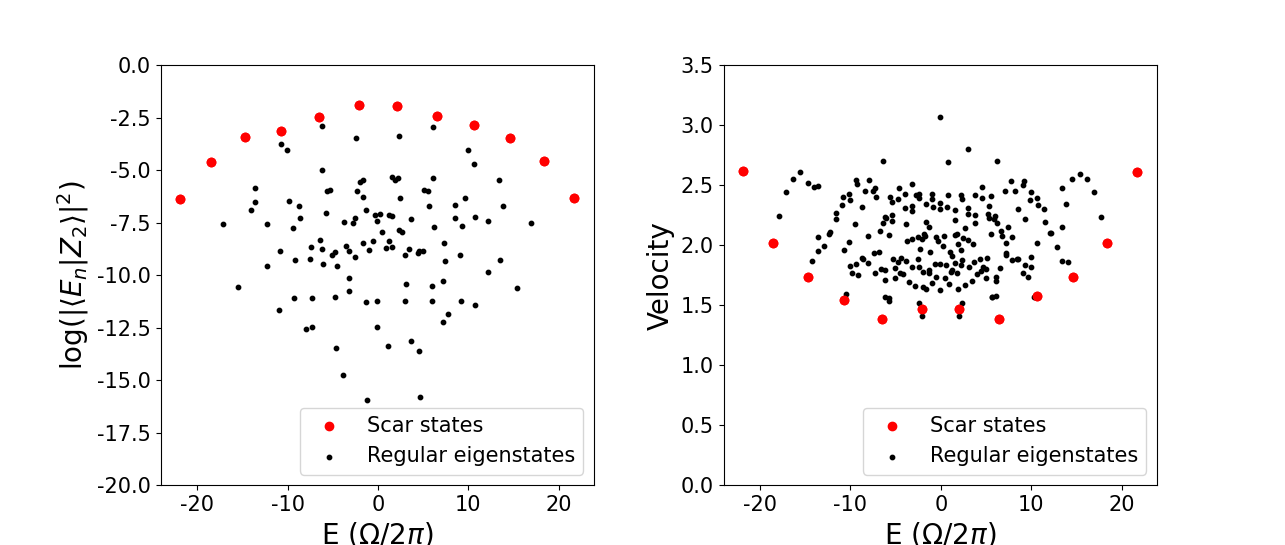}
\caption{\label{fig:eigenstate_velocity} Distribution of eigenstate properties for the Rydberg Hamiltonian Eq.~\ref{eq:rydberg}. (a) Overlap $|\langle E_n|{\bf Z}_2\rangle|^2$ between eigenstates $|E_n\rangle$ and the $\left|{\bf Z}_2\right\rangle$ state, showing strong overlap with quantum many-body scar eigenstates (red markers) with $\left|{\bf Z}_2\right\rangle$. (b) Butterfly velocities extracted from OTOC fitting, demonstrating anomalously slow information spreading for quantum many-body scar eigenstates. Numerical results for $N=11$ spin chain with open boundary conditions.}
\end{figure}

In this section, we use the eigenstates of the system as initial states. Figure \ref{fig:eigenstate_velocity} (a) displays the energy of each eigenstate and the overlap between the $\left|{\bf Z}_2\right\rangle$ state
with the eigenstates. The red dots in the figure represent eigenstates within the scar subspace. Figure~\ref{fig:eigenstate_velocity} (b) shows the fitted wavefront propagation velocity of their corresponding OTOC. The eigenstates in the scar subspace exhibit significantly slower butterfly velocities compared to other states. The substantial overlap of $\left|{\bf Z}_2\right\rangle$ state with the eigenstates in the scar subspace enable us to dynamically enter scar subspace and experimentally verify the actual slowing down of the butterfly velocity. In contrast, the $\left|\bold{g}\right>$ state has far less overlap with the eigenstates in the scar subspace, representing states outside the scar subspace.

\subsection{OTOCs for product states}

\begin{figure}[!htp]
    \includegraphics[width=0.65\linewidth]{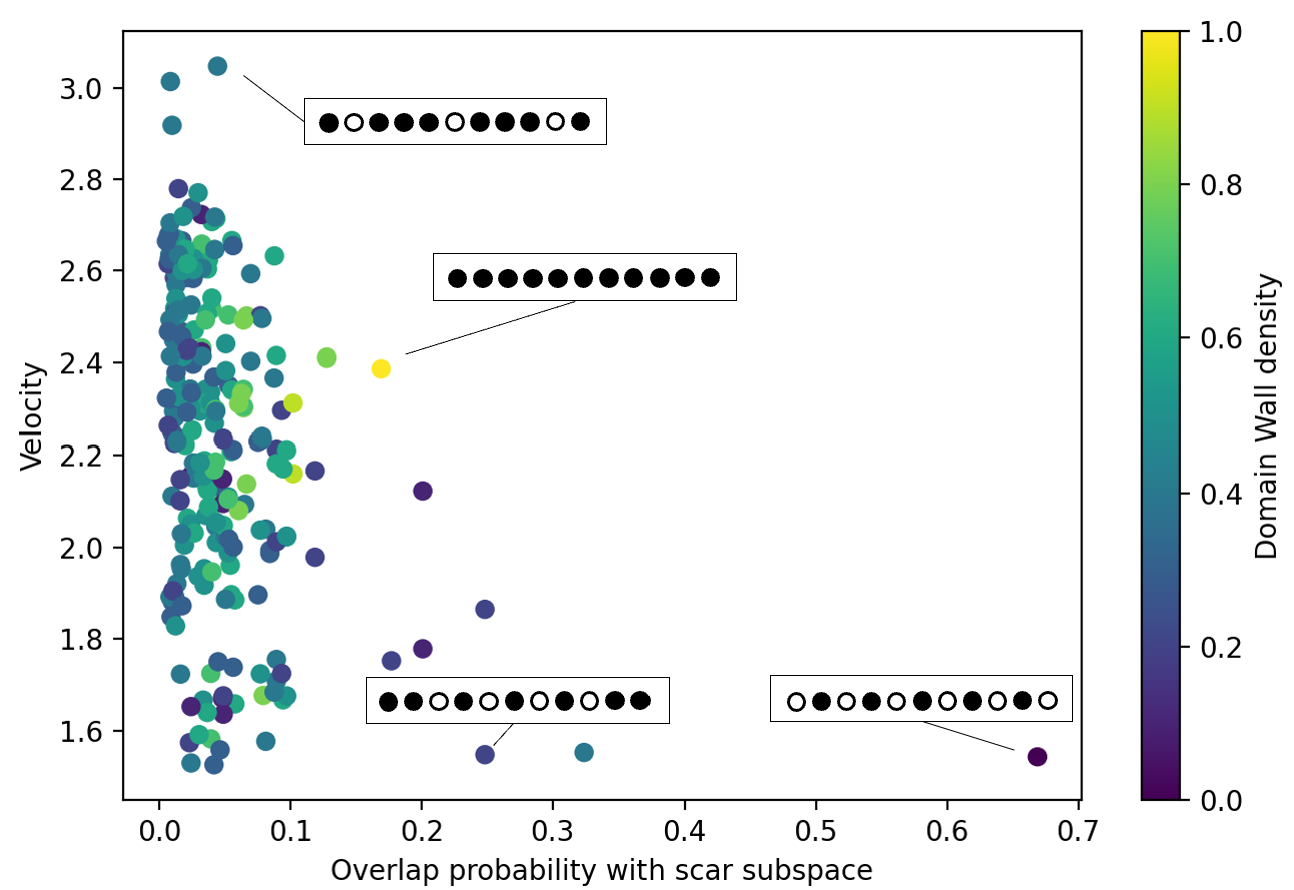}
    \caption{\label{fig:velocityvsoverlap} OTOCs for product states. Each point in the scatter plot represents an initial state in the form of a product state allowed in the PXP model. The horizontal axis shows the probability of the product state in the scar subspace, while the vertical axis represents the corresponding fitted information propagation velocity. The $\left|{\bf Z}_2\right\rangle$ state in the lower right corner represents the product state that can be experimentally prepared exhibiting the largest overlap with the scar subspace and the slowest propagation velocity. In the lower left corner, other product states similar to the $\left|{\bf Z}_2\right\rangle$ state, also exhibit slower propagation velocities. The $\left|\bold{g}\right>$ state represents product states with major components outside the scar subspace, showing faster fitted information propagation velocities.}

\end{figure}

We also employ product states that are easily prepared in experiments for numerical simulations. Different product states exhibit varying degrees of overlap with the scar subspace. Figure \ref{fig:velocityvsoverlap} shows the overlap probabilities of each state in the scar subspace and the distribution of fitted OTOC wavefront propagation velocities. The $\left|{\bf Z}_2\right\rangle$ state and $\left|\bold{g}\right>$ state are shown to be typical representatives of these states.

\subsection{OTOCs for product states with Rydberg clusters}

\begin{figure}[h]
    \includegraphics[width=\linewidth]{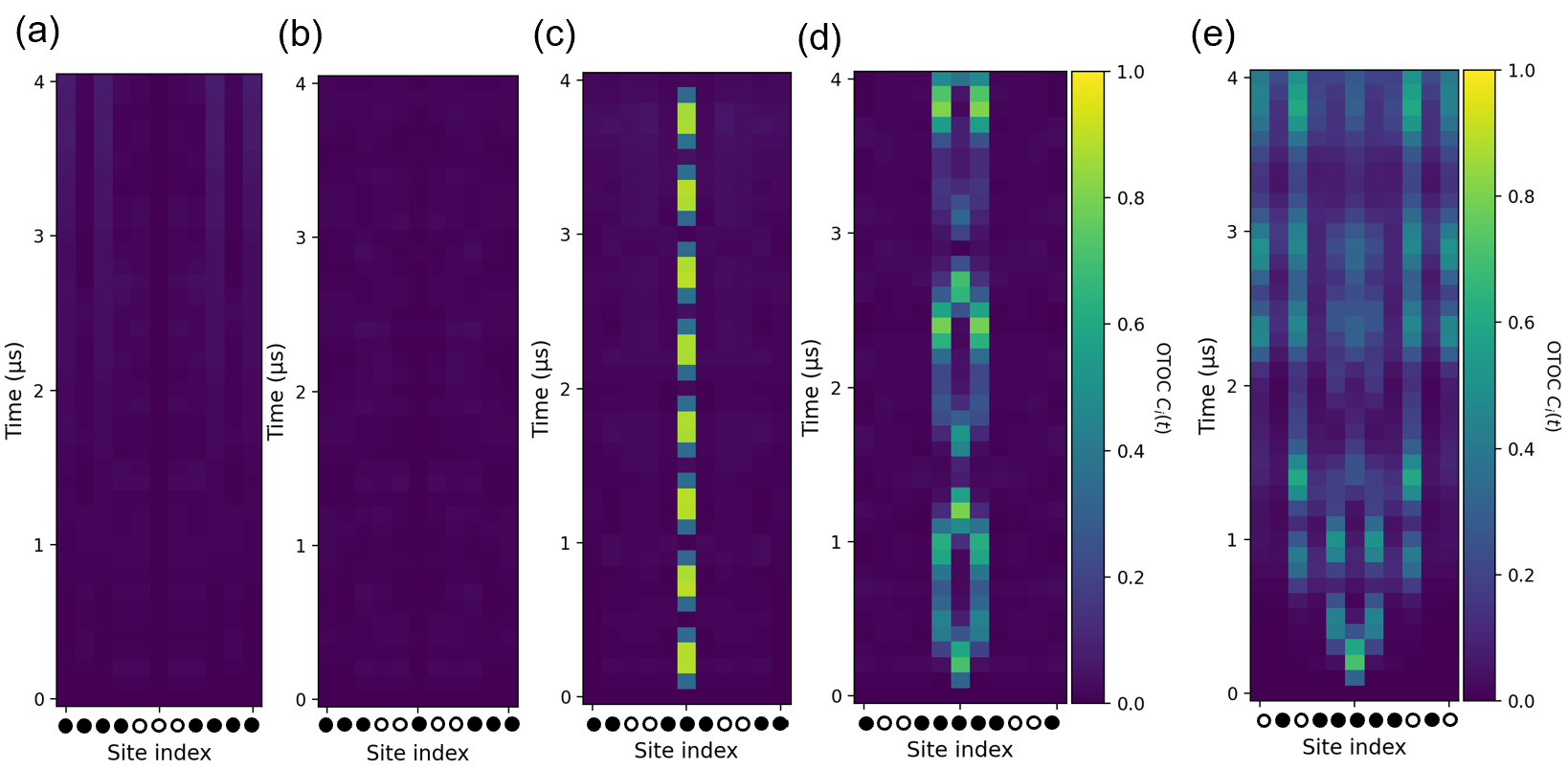}
    \caption{\label{fig:rydbergclaster} Freezing or trapping of OTOCs with initial states containing Rydberg clusters. For initial states with two or more NN Rydberg excitations, the OTOC signal is suppressed to nearly vanishing values (a) within the Rydberg cluster and (b) at sites adjacent to the cluster. Information propagation from neighboring sites is inhibited and reflected at the cluster, resulting in (c-d) trapped information between two Rydberg clusters, which is distinctly different from (e) the initial state without a Rydberg cluster.}
    \end{figure}

For initial states not captured by the PXP Hamiltonian but realizable in the Rydberg system, certain special states exhibit information freezing and trapping phenomena, which is related to the Hilbert space fragmentation. As illustrated in the Fig. \ref{fig:rydbergclaster}, for initial states with two or more NN Rydberg excitations, information originating from such Rydberg clusters remains frozen, with the OTOC exhibiting near-zero values. Meanwhile, information propagating from adjacent sites is blocked and reflected at the cluster locations. When local information is confined between two Rydberg clusters, it becomes trapped in a localized region, exhibiting near-zero propagation speed over extended time periods.



\bibliographystyle{apsrev4-2}
\bibliography{sup}